\documentclass[10pt]{amsart}
\usepackage{amsmath,amsfonts,amscd,amssymb,epsf}
\newtheorem{theorem}{Theorem}[section]
\newtheorem{proposition}[theorem]{Proposition}

\theoremstyle{definition}

\theoremstyle{remark} 
\numberwithin{equation}{section}

\newcommand{\Z}{{\mathbb{Z}}}\newcommand{\Del}{\mathbb{D}}

\newcommand{\C}{{\mathbb{C}}}
\newcommand{\R}{{\mathbb{R}}}
\newcommand{\mL}{\mathcal{L}}

\newcommand{\pa}{\partial}

\begin{document}
\title[Conformal Mappings and Dispersionless Toda hierarchy II]{Conformal Mappings and Dispersionless Toda hierarchy II: General String Equations}
\author{ Lee-Peng Teo}   \email{lpteo03@yahoo.com}

\begin{abstract}In this article, we classify the solutions of the dispersionless Toda hierarchy into degenerate and non-degenerate cases. We show that every non-degenerate solution is determined by a function $\mathcal{H}(z_1,z_2)$ of two variables. We interpret these non-degenerate solutions as defining evolutions on the space $\mathfrak{D}$ of pairs of conformal mappings $(g,f)$, where $g$ is a univalent function on the exterior of the unit disc, $f$ is a univalent function on the unit disc, normalized such that $g(\infty)=\infty$, $f(0)=0$ and $f'(0)g'(\infty)=1$. For each solution, we show how to define the natural time variables $t_n, n\in\Z$, as complex coordinates on the space $\mathfrak{D}$. We also find explicit formulas for the tau function of the dispersionless Toda hierarchy in terms of $\mathcal{H}(z_1, z_2)$. Imposing some conditions on the function $\mathcal{H}(z_1, z_2)$, we show that the dispersionless Toda flows can be naturally restricted to the subspace $\Sigma$ of $\mathfrak{D}$ defined by $f(w)=1/\overline{g(1/\bar{w})}$. This recovers the result of Zabrodin   \cite{31}.
\end{abstract}
 \maketitle

\section{Introduction}

This paper is a continuation of our previous work \cite{1}, where we considered    evolutions of conformal mappings  described by an infinite hierarchy of dispersionless Toda flows \cite{2,3} which satisfies the string equation. In this paper, we are going to consider the general evolutions of conformal mappings that can be described by dispersionless Toda hierarchies and derive the corresponding  string equations.

Dispersionless Toda hierarchy was introduced in \cite{2,3} as dispersionless limit of the Toda lattice hierarchy \cite{5}. It describes the evolutions of two formal power series \begin{equation}\label{eq6_5_6}\begin{split}\mathcal{L}(w) =& r(\boldsymbol{t}) w+ \sum_{n=0}^{\infty}u_{n+1}(\boldsymbol{t})w^{-n}, \hspace{0.5cm}\tilde{\mathcal{L}}(w)^{-1}=r(\boldsymbol{t})w^{-1}+\sum_{n=0}^{\infty}\tilde{u}_{n+1}(\boldsymbol{t})w^{n}, \end{split}\end{equation} with respect to an infinite set of time variables $t_n, n\in \Z$, denoted collectively by $\boldsymbol{t}$, by the following Lax equations:
\begin{equation}\label{eq6_5_4}
\begin{split}
\frac{\pa\mathcal{L}}{\pa t_n} =\left\{ \mathcal{B}_{n}, \mathcal{L}\right\}, \hspace{1cm}
\frac{\pa\tilde{\mathcal{L}}}{\pa t_n} =\left\{ \mathcal{B}_{n}, \tilde{\mathcal{L}}\right\}.
\end{split}
\end{equation}
 Here
$\mathcal{B}_n$ is defined by
\begin{equation*}\begin{split}
\mathcal{B}_n(w) =& \left(\mathcal{L}(w)^n\right)_{>0}+\frac{1}{2}\left(\mathcal{L}(w)^n\right)_{0}, \;\; n\geq 1, \hspace{1cm} \mathcal{B}_0(w)=\log w,  \\\mathcal{B}_{n}(w) = &\left(\tilde{\mathcal{L}}(w)^{n}\right)_{<0}+\frac{1}{2}\left(\tilde{\mathcal{L}}(w)^{n}\right)_{0}, \;\;n\leq -1,\end{split}
\end{equation*}  and the Poisson bracket is defined by
\begin{equation*}
\left\{ F_1(w), F_2(w)\right\} =w\frac{\pa F_1(w)}{\pa w}\frac{\pa F_2(w)}{\pa t_0} -w\frac{\pa F_2(w)}{\pa w}\frac{\pa F_1(w)}{\pa t_0}.
\end{equation*}

As mentioned in our previous paper \cite{1}, the integrable structures of conformal mappings have attracted considerable interest since the work of Wiegmann and Zabrodin \cite{4}. Given a simple analytic curve $\mathcal{C}$ that separates $0$ and $\infty$ on the extended complex plane $\hat{\C}$, let $\Omega^+$ be the interior domain that contains the origin and $\Omega^-$ the exterior domain. Denote by $g(w)$ the unique conformal map  mapping the exterior of the unit disc $\Del^*$ to the exterior domain which satisfies the normalization conditions $g(\infty)=\infty$ and $g'(\infty)>0$. Wiegmann and Zabrodin \cite{4} defined the time variables $t_n, n\geq 1$, in terms of the harmonic moments of the exterior domain $\Omega^-$:
\begin{equation*}\begin{split}
t_n =-\frac{1}{\pi n}\iint_{\Omega^-} z^{-n}d^2z = \frac{1}{2\pi in}\oint_{\mathcal{C}}z^{-n} \bar{z} dz,
\end{split}
\end{equation*}and defined $t_0$ in terms of the area of the interior domain:
\begin{equation*}
\begin{split}
t_0=\frac{1}{\pi}\iint_{\Omega^+} d^2z =\frac{1}{2\pi i}\oint_{\mathcal{C}}\bar{z} dz.
\end{split}
\end{equation*}They showed that by taking $t_{-n}=-\bar{t}_n$ for  $n\geq 1$,  the evolutions of $(\mathcal{L}(w)=g(w), \tilde{\mathcal{L}}(w)=1/\overline{g(1/\bar{w})})$ with respect to $t_n, n\in\Z,$ satisfy the dispersionless Toda hierarchy. This solution of the dispersionless Toda hierarchy satisfies the string equation
\begin{equation}\label{eq6_4_1}
\begin{split}
\left\{ \mathcal{L}(w), \tilde{\mathcal{L}}(w)^{-1}\right\} =1.
\end{split}
\end{equation} A tau function $\tau(\boldsymbol{t})$ was also constructed and shown to generate the harmonic moments of the interior domain $\Omega^+$. More precisely, it was proved that for $n\geq 1$,
\begin{equation*}
\frac{\pa \log\tau}{\pa t_n} = \frac{1}{\pi} \iint_{\Omega_+} z^n d^2z= \frac{1}{2\pi i}\oint_{\mathcal{C}}z^n \bar{z} dz.
\end{equation*}
This work of Wiegmann and Zabrodin has later been reinterpreted and elaborated from different perspectives such as tau function and inverse potential problem \cite{14}, interface dynamics, Laplacian growth or Hele-Shaw flow problem \cite{6, 13, 12, 19, 24, 26, 28,23,22}, Dirichlet boundary value problem \cite{9, 29}, quantum field theory of free bosons \cite{7}, large $N$-limit of normal matrix ensemble \cite{15,11, 16} and string equation and string theory   \cite{17, 25, 27}. The extension of these to multiply connected domains have been considered in \cite{18, 10, 20, 21}.

  As an integrable hierarchy, the dispersionless Toda hierarchy was proposed in the real domain where all the time variables and the the coefficients $r, u_n, \tilde{u}_n, n\geq 1$, of $\mathcal{L}$ and $\tilde{\mathcal{L}}$ are real-valued. Moreover, $t_n, n\in\Z,$ are independent variables.  However, as one can see from the discussion above, for the solution provided by Wiegmann and Zabrodin, the time variables $t_n, n\in \Z$, are in general complex-valued, so are the coefficients $u_n, \tilde{u}_n, n\geq 1$. Moreover, the time variables are independent over $\R$, but are not independent over $\C$, since for $n\geq 1$, $t_{-n}$ and $t_n$ are related by $t_{-n}=-\bar{t}_n$. This leads to the relation $\tilde{\mathcal{L}}^{-1}(w) = \overline{\mathcal{L}(1/\bar{w})}$ of the two power series $\mathcal{L}(w)$ and $\tilde{\mathcal{L}}(w)$. Therefore, for this solution, only half of the flows are independent (over $\C$). In other words, it can be considered as a solution of a reduction of the dispersionless Toda hierarchy. In our previous work \cite{1}, we extended the work of Wiegmann and Zabrodin by considering pairs of conformal mappings $(g, f)$, where $g$ is conformal on the exterior of the unit disc $\Del^*$ normalized such that $g(\infty)=\infty$, and $f$ is conformal on the unit disc normalized such that $f(0)=0$. $f$ and $g$ are only required to be related by $f'(0)g'(\infty)=1$. We showed that by suitably defining $t_n, n\in \Z$, which are in general complex-valued and independent over $\C$, the dynamics of the pair of conformal maps $(g,f)$ is described by the dispersionless Toda flows, but with $t_n, n\in \mathbb{Z}$, and $r, u_n, \tilde{u}_n, n\in \mathbb{N}$, complex-valued, and $r, u_n, \tilde{u}_{n}, n\in \mathbb{N}$, are holomorphic in $t_n, n\in \mathbb{Z}$. We also constructed a tau function for the hierarchy which is real-valued. By restricting our flows to the subspace $t_{-n}+\bar{t}_n=0$ for $n\geq 1$, and $t_0=\bar{t}_0$, we recovered the flows considered by Wiegmann and Zabrodin. On the other hand, by restricting our flows to the subspace where $t_n=\bar{t}_n, n\in\Z$,   or equivalently where $f$ and $g$ have real coefficients, we obtained the usual solution of the dispersionless Toda hierarchy where all the variables are real. The tau function we constructed can then be interpreted as the free energy of a two Hermitian matrix model \cite{30}.
  From a different perspective, we have considered a particular solution of the complexified version of the dispersionless Toda hierarchy which satisfies the string equation \eqref{eq6_4_1} and interpreted it as describing evolutions of pairs of conformal mappings.

 Since the works on evolutions of conformal mappings have found applications in a lots of different areas, it is natural to ask what are the general solutions of dispersionless   Toda hierarchies and whether they can be interpreted as evolutions of conformal mappings. In fact, soon after the work \cite{4}, Zabrodin \cite{31} has shown that one can obtain  $\left(\mathcal{L}(w)=g(w), \tilde{\mathcal{L}}(w)=1/\overline{g(1/\bar{w})}\right)$ as more general solutions of dispersionless Toda hierarchy which satisfy the generalized string equation
\begin{equation}\label{eq6_4_1_1}
\begin{split}
\left\{ \mathcal{L}(w), \tilde{\mathcal{L}}(w)^{-1}\right\} =\frac{1}{\pa_z\pa_{\bar{z}}\mathcal{U}(\mL(w),\tilde{\mL}(w^{-1})^{-1})},
\end{split}
\end{equation}  where $\mathcal{U}(z,\bar{z})$ is a real-valued function of $z$ and $\bar{z}$, by defining the time variables $t_n, n\geq 0$, as
\begin{equation*}
\begin{split}
t_n =\frac{1}{2\pi i n}\oint_{\mathcal{C}} z^{-n} \pa_z\mathcal{U}(z, \bar{z}) dz, \;\; n\geq 1,\hspace{1cm} t_0 =\frac{1}{2\pi i }\oint_{\mathcal{C}}  \pa_z\mathcal{U}(z, \bar{z}) dz,
\end{split}
\end{equation*} and let $t_{-n}=-\bar{t}_n$ for $n\geq 1$.  A tau function for the problem was also derived by using electrostatic variational principle. The particular case considered in \cite{4} corresponds to choosing $\mathcal{U}(z,\bar{z})=z\bar{z}$. As in \cite{4}, the  solutions  provided by Zabrodin \cite{31} should be considered as solutions to reductions of complexified  dispersionless Toda hierarchy characterized by $t_{-n}+\bar{t}_n=0$ for $n\geq 1$ and $t_0=\bar{t}_0$.

  In this paper, we consider general solutions of complexified dispersionless Toda hierarchy where $r, u_n, \tilde{u}_n, n\in\mathbb{N}$, depend holomorphically on $t_m, m\in\mathbb{Z}$. We show that these solutions can be interpreted as describing evolutions of pairs of conformal mappings $(g,f)$ when one defines the time variables $t_n, n\in \Z$, appropriately. We also construct a real-valued tau function for each solution of the dispersionless Toda hierarchy. Under certain reality conditions, we show that the solutions of Zabrodin \cite{31} can be considered as the restriction of our solution to the subspace defined by $t_{-n}+\bar{t}_n=0, n\geq 1,$ and $t_0=\bar{t}_0$.    From the perspective of dispersionless Toda hierarchy, this work answers the question: What is the general solution of dispersionless Toda hierarchy and what is the corresponding tau function? From the perspective of conformal mappings, this work characterizes all different complex coordinates on the space of pairs of conformal mappings which can give rise to dispersionless Toda flows.

\section{Generalized Grunsky coefficients and Faber polynomials}\label{sec2}To make this paper self-contained, we review again the concepts of generalized  Grunsky coefficients and Faber polynomials here.
Let $F(z) = \alpha_1z +\alpha_2 z^2 + \ldots$ be a function
univalent in a neighborhood of the origin and $G(z) =
\beta z + \beta_0 + \beta_1 z^{-1} + \ldots$  be a function
univalent in a neighborhood of $\infty$ such that $\alpha_1\beta=1$. We define the
generalized Grunsky coefficients $b_{m,n}$, $m,n\in\Z$, and Faber
polynomials $P_n(z)$, $n\in \Z$, of the pair $(G, F)$ by the following formal power series
expansion:
\begin{equation}\begin{split}\label{eq6_5_1}
\log \frac{G(z)-G(\zeta)}{z-\zeta}&=
b_{0,0}-\sum_{m=1}^{\infty}\sum_{n=1}^{\infty} b_{mn} z^{-m}\zeta^{-n},\\
\log \frac{G(z) - F(\zeta)}{z} &= b_{0,0}-\sum_{m=1}^{\infty}
\sum_{n=0}^{\infty} b_{m,-n} z^{-m} \zeta^n,\\
\log \frac{F(z)-F(\zeta)}{z-\zeta} &=
-\sum_{m=0}^{\infty}
\sum_{n=0}^{\infty} b_{-m, -n} z^{m} \zeta^n,\\
\log \frac{G(z) -w}{\beta z} &= -\sum_{n=1}^{\infty}
\frac{P_n(w)}{n}
z^{-n} ,\\
\log \frac{w- F(z)}{w} &= \log
\frac{F(z)}{\alpha_1 z}-\sum_{n=1}^{\infty} \frac{P_{-n}(w)}{n}
z^{n}.\end{split}
\end{equation}
  For $m\geq 0$, $n\geq 1$, $b_{-m,n}:=b_{n,-m}$, and by convention, $P_0(w):=\log w$ (which is not a polynomial). By definition,
the Grunsky coefficients are symmetric, i.e., $b_{m,n} =b_{n,m}$
for all $m,n\in\Z$. The coefficient $b_{0,0}$ is given explicitly by $-\log \alpha_1=\log \beta$, where $\alpha_1= F'(0)$ and $\beta=G'(\infty)$. For $n\geq 1$, $P_n(w)$ is a polynomial of degree $n$ in $w$
and $P_{-n}(w)$ is a polynomial of degree $n$ in $1/w$, which can be defined alternatively by
\begin{equation*}\begin{split}
P_n(w) = (G^{-1}(w)^n)_{\geq 0}, \hspace{1cm}P_{-n}(w) =
\left( F^{-1}(w)^{-n}\right)_{\leq 0}.\end{split}\end{equation*} Here when $S$ is a subset
of integers and $A(w)=\sum_{n} A_nw^n$ is a (formal) power series,
we denote by $(A(w))_{S}$ the truncated sum $\sum_{n\in S} A_n w^n$.

From \eqref{eq6_5_1}, we can deduce the following:
\begin{equation}\label{eq6_5_2}\begin{split}
\log\frac{G(z)}{z} &= b_{0,0}-\sum_{m=1}^{\infty}
b_{0,m} z^{-m},\hspace{1.5cm} \log\frac{F(z)}{z} = -\sum_{m=0}^{\infty}
b_{0,-m} z^{m},\end{split}
\end{equation}and for $n\geq 1$,\begin{equation}\label{eq6_5_3}\begin{split}
P_n (G(z)) &= z^n + n\sum_{m=1}^{\infty} b_{nm}
z^{-m},\hspace{1.5cm}P_n (F(z)) =
nb_{n,0}+n\sum_{m=1}^{\infty}
b_{n, -m} z^m,  \\
P_{-n}(G(z)) &= -nb_{-n,0} + n\sum_{m=1}^{\infty}
b_{-n,m} z^{-m},\hspace{0.4cm}P_{-n} (F(z))=
z^{-n} + n\sum_{m=1}^{\infty} b_{-n,-m} z^{m}.\end{split}
\end{equation}

It follows that
\begin{equation}\label{eq6_8_4}
\begin{split}
P_0'(G(z))G'(z)&=\frac{1}{z} +\sum_{m=1}^{\infty}m
b_{0,m} z^{-m-1},\hspace{1cm} P_0'(F(z))F'(z) =\frac{1}{z}-\sum_{m=1}^{\infty}
m b_{0,-m} z^{m-1},\\
P_n' (G(z))G'(z) &= nz^{n-1} - n\sum_{m=1}^{\infty}m b_{nm}
z^{-m-1},\hspace{0.5cm}P_n' (F(z))F'(z) =
 n\sum_{m=1}^{\infty}m
b_{n, -m} z^{m-1},  \\
P_{-n}'(G(z)) G'(z)&=  - n\sum_{m=1}^{\infty}m
b_{-n,m} z^{-m-1},\hspace{0.4cm}P_{-n}' (F(z))F'(z)=
-nz^{-n-1} + n\sum_{m=1}^{\infty}m b_{-n,-m} z^{m-1}.
\end{split}
\end{equation}

\section{Dispersionless Toda hierarchy and its general solutions}

As discussed in the introduction, dispersionless Toda hierarchy is a hierarchy of equations which can be put into the Lax form \eqref{eq6_5_4}. In this section, we first review some basic facts we need later. We then classify and characterize the solutions of the dispersionless Toda hierarchy.

\subsection{Orlov-Schulman functions}
First, recall that if $(\mathcal{L}, \tilde{\mathcal{L}})$ are power series of the form \eqref{eq6_5_6} that satisfy the dispersionless Toda hierarchy \eqref{eq6_5_4}, their Orlov-Schulman functions are functions $\mathcal{M}(w)$ and $\tilde{\mathcal{M}}(w)$ of the form
\begin{equation}\label{eq6_5_7}
\begin{split}
 \mathcal{M}(w) =&\sum_{n=1}^{\infty} n t_n\mathcal{L}(w)^n + t_0 +\sum_{n=1}^{\infty} v_n \mathcal{L}(w)^{-n},
 \\
 \tilde{\mathcal{M}}(w) =&-\sum_{n=1}^{\infty}nt_{-n} \tilde{\mathcal{L}}(w)^{-n} + t_0 -\sum_{n=1}^{\infty} v_{-n}\tilde{ \mathcal{L}}(w)^{n}
\end{split}
\end{equation}that satisfy the Lax equations
\begin{equation}\label{eq6_5_9}
\frac{\pa\mathcal{M}}{\pa t_n} =\left\{ \mathcal{B}_{n}, \mathcal{M}\right\}, \hspace{1cm}
\frac{\pa\tilde{\mathcal{M}}}{\pa t_n} =\left\{ \mathcal{B}_{n}, \tilde{\mathcal{M}}\right\}, \;\;\;\;n\in\Z,
\end{equation}and the canonical Poisson relations
\begin{equation}\label{eq6_5_8}\begin{split}
\left\{\mathcal{L}, \mathcal{M}\right\}=\mathcal{L}, \hspace{1cm}\left\{\tilde{\mathcal{L}}, \tilde{\mathcal{M}}\right\}=\tilde{\mathcal{L}}.
\end{split}
\end{equation}

\subsection{Phi function $\phi$ and tau function $\tau$}
Given a solution  $(\mathcal{L}, \tilde{\mathcal{L}})$ of the dispersionless Toda hierarchy \eqref{eq6_5_4}, there exists a phi function $\phi(\boldsymbol{t})$ and a tau function $\tau(\boldsymbol{t})$ such that
\begin{equation*}
\frac{\pa\phi}{\pa t_n} =\frac{\pa v_n}{\pa t_0} \hspace{1cm}\text{and}\hspace{1cm}\frac{\pa\log\tau}{\pa t_n}= v_n\hspace{1cm}\text{for all}\;\; n\in\Z.
\end{equation*}Here $v_0:=\phi$. If we let $G(z)$ and $F(z)$ be formally the inverses of $\mathcal{L}(w)$ and $\tilde{\mathcal{L}}(w)$ respectively, i.e., $\mathcal{L}(G(z))=z$ and $\tilde{\mathcal{L}}(F(z))=z$, and let $b_{m,n}, m,n\in\Z,$ be the generalized Grunsky coefficients of the pair $(G, F)$, then
\begin{equation}\label{eq6_5_10}
\frac{\pa^2\log \tau(\boldsymbol{t})}{\pa t_m \pa t_n}=\begin{cases}-|mn|b_{m,n}(\boldsymbol{t}), \hspace{1cm}&\text{if}\;\; mn\neq 0,\\
|m|b_{m,0}(\boldsymbol{t}), &\text{if}\;\; m\neq 0, n=0,\\-2b_{0,0}(\boldsymbol{t}), &\text{if}\;\; m=n=0.\end{cases}
\end{equation}Conversely, if $b_{m,n}, m,n\in\Z,$ are the generalized Grunsky coefficients of the pair $(G, F)$, and $\tau(\boldsymbol{t})$ is a  function satisfying
\eqref{eq6_5_10}, then $(\mathcal{L}(w)=G^{-1}(w), \tilde{\mathcal{L}}(w)=F^{-1}(w))$ is a solution of the dispersionless Toda hierarchy.

\subsection{Riemann-Hilbert data}

The Riemann-Hilbert data of a solution $(\mathcal{L}, \tilde{\mathcal{L}})$ of the dispersionless Toda hierarchy is a pair of functions $U(w,t_0)$ and $V(w,t_0)$ of the variables $w$ and $t_0$ that satisfy
\begin{equation}\label{eq6_5_11}
\tilde{\mathcal{L}}= U(\mathcal{L},\mathcal{M}), \hspace{1cm} \tilde{\mathcal{M}}= V(\mathcal{L},\mathcal{M}),
\end{equation}and the canonical Poisson relation
\begin{equation}\label{eq6_5_12}
\left\{U,V\right\} = w\frac{\pa U}{\pa w}\frac{\pa V}{\pa t_0}-w\frac{\pa V}{\pa w}\frac{\pa U}{\pa t_0}=U.
\end{equation}It was shown in \cite{3} that there always exists a Riemann-Hilbert data for any solutions of the dispersionless Toda hierarchy. Conversely, it was also proved that  if $U(w,t_0)$ and $V(w,t_0)$ are functions satisfying the canonical Poisson relation \eqref{eq6_5_12}, and $\mathcal{L}, \tilde{\mathcal{L}}$ are functions of the form \eqref{eq6_5_6},   $\mathcal{M}, \tilde{\mathcal{M}}$ are functions of the form \eqref{eq6_5_7}, and they satisfy \eqref{eq6_5_11}, then $(\mathcal{L}, \tilde{\mathcal{L}})$ is a solution of the dispersionless Toda hierarchy with Orlov-Schulman functions $\mathcal{M}(w)$ and $\tilde{\mathcal{M}}(w)$.

\subsection{General solutions of dispersionless Toda hierarchy}
Now we come to the classification and characterization of the solutions of the dispersionless Toda hierarchy.
Suppose $(\mathcal{L}, \tilde{\mathcal{L}})$  is a solution of the dispersionless Toda hierarchy with Orlov-Schulman functions $\mathcal{M}(w)$, $\tilde{\mathcal{M}}(w)$
and Riemann-Hilbert data $U(w,t_0)$ and $V(w,t_0)$. We have two cases. First, if $U$ is independent of $t_0$, i.e., $\pa U/\pa t_0=0$,  then the canonical Poisson relation \eqref{eq6_5_12} implies that $U'(w)\neq 0$ and
\begin{equation*}
V(w, t_0)= \frac{U(w)}{wU'(w)}t_0+ U_1(w),
\end{equation*}for an arbitrary function $U_1(w)$. In this case, we find that
\begin{equation*}
\tilde{\mathcal{L}}=U(\mathcal{L}), \hspace{1cm} \tilde{\mathcal{M}}= \frac{U(\mathcal{L})}{\mathcal{L}U'(\mathcal{L})}\mathcal{M}+U_1(\mathcal{L}),
\end{equation*}  and
\begin{equation*}
\left\{ \mathcal{L}, \tilde{\mathcal{L}}\right\}=0.
\end{equation*}
This should be considered as a degenerate solution of the dispersionless Toda hierarchy since the relation $\tilde{\mathcal{L}}=U(\mathcal{L})$ implies that for all $n\in\Z$, the set of equations $$\frac{\pa\tilde{\mathcal{L}}}{\pa t_n}=\left\{\mathcal{B}_n, \tilde{\mathcal{L}}\right\}$$ follows immediately from the set of equations$$\frac{\pa \mathcal{L}}{\pa t_n}=\left\{\mathcal{B}_n, \mathcal{L}\right\}.$$

In the second case,  $\pa U/\pa t_0$ does not vanish. Then inverse function theorem implies that we can solve $t_0$ as a function of $w$ and $\tilde{w}$ from $\tilde{w}=U(w,t_0)$. More precisely, there exists a function   $\mathcal{A}(w,\tilde{w})$ such that
\begin{equation*}
\tilde{w} = U\left(w, \mathcal{A}(w, \tilde{w})\right).
\end{equation*}Moreover,
\begin{equation*}
\frac{\pa \mathcal{A}}{\pa w}(w, \tilde{w})=- \frac{\frac{\pa U}{\pa w}\left(w, \mathcal{A}(w, \tilde{w})\right)}{\frac{\pa U}{\pa t_0}\left(w, \mathcal{A}(w, \tilde{w})\right)},\hspace{1cm}\frac{\pa \mathcal{A}}{\pa \tilde{w}}(w, \tilde{w})= \frac{1}{\frac{\pa U}{\pa t_0}\left(w, \mathcal{A}(w, \tilde{w})\right)}.
\end{equation*}Obviously, $\pa\mathcal{A}/\pa \tilde{w}\neq 0$. Define the function $\tilde{\mathcal{A}}(w,\tilde{w})$ by
\begin{equation*}
\tilde{\mathcal{A}}(w,\tilde{w})=V\left(w, \mathcal{A}(w, \tilde{w})\right).
\end{equation*}Then from the canonical Poisson relation \eqref{eq6_5_12}, we have
\begin{equation*}\begin{split}
\frac{\pa\tilde{\mathcal{A}}}{\pa w} =& \frac{\pa V}{\pa w}+\frac{\pa V}{\pa t_0}\frac{\pa \mathcal{A}}{\pa w}=\frac{1}{\frac{\pa U}{\pa t_0}}\left(\frac{\pa V}{\pa w}\frac{\pa U}{\pa t_0} -\frac{\pa U}{\pa w}\frac{\pa V}{\pa t_0}\right)=-\frac{ \tilde{w}  }{w}\frac{\pa \mathcal{A}}{\pa \tilde{w}}(w,\tilde{w}).
\end{split}
\end{equation*}This gives
\begin{equation*}
\frac{\pa}{\pa w}\left( -\frac{ \tilde{\mathcal{A}}}{\tilde{w}} \right)=\frac{\pa }{\pa \tilde{w}}\left(\frac{\mathcal{A}}{w}\right).
\end{equation*}Consequently, there exists a function $\mathcal{H}\left(w,\tilde{w}\right)$ such that
\begin{equation}\label{eq6_5_15}
\mathcal{A}\left(w, \tilde{w}\right) =w\pa_w\mathcal{H}\left(w,\tilde{w}\right), \hspace{1cm}\tilde{\mathcal{A}}\left(w, \tilde{w}\right) =-\tilde{w}\pa_{\tilde{w}}\mathcal{H}\left(w,\tilde{w}\right).
\end{equation}The condition $\pa\mathcal{A}/\pa \tilde{w}\neq 0$ is then equivalent to \begin{equation}\label{eq6_5_13}\frac{\pa^2 \mathcal{H}}{\pa w\pa\tilde{w}}\neq 0.\end{equation}The relations \eqref{eq6_5_11} become
\begin{equation}\label{eq6_5_14}
\mathcal{M}= \mathcal{L}\pa_{\mathcal{L}}\mathcal{H}\left(\mathcal{L}, \tilde{\mathcal{L}}\right), \hspace{1cm}\tilde{\mathcal{M}}= -\tilde{\mathcal{L}}\pa_{\tilde{\mathcal{L}}}\mathcal{H}\left(\mathcal{L}, \tilde{\mathcal{L}}\right).
\end{equation} Moreover,
\begin{equation*}
\left\{\mathcal{L}, \tilde{\mathcal{L}}\right\} = \mathcal{L}\frac{\pa U}{\pa t_0}\left( \mathcal{L}, \mathcal{A}\left(\mathcal{L},\tilde{\mathcal{L}}\right)\right)=\frac{1}{\pa_{\mathcal{L}}\pa_{\tilde{\mathcal{L}}}\mathcal{H}(\mathcal{L}, \tilde{\mathcal{L}})},
\end{equation*}which is the string equation for this solution.

Conversely, suppose $\mathcal{H}(w,\tilde{w})$ is a function satisfying \eqref{eq6_5_13}, and $\mathcal{L}, \tilde{\mathcal{L}}$ are functions of the form \eqref{eq6_5_6},   $\mathcal{M}, \tilde{\mathcal{M}}$ are functions of the form \eqref{eq6_5_7}, so that \eqref{eq6_5_14} holds. Define the functions $\mathcal{A}(w,\tilde{w})$ and $\tilde{\mathcal{A}}(w,\tilde{w})$ by the relations \eqref{eq6_5_15}. Then $\pa\mathcal{A}/\pa \tilde{w}\neq 0$. Therefore, we can solve $\tilde{w}$ as a function of $w$ and $t_0$ from the equation $$t_0=\mathcal{A}(w, \tilde{w}),$$ which we denote by $U(w, t_0)$, so that
\begin{equation*}
t_0=\mathcal{A}(w, U(w, t_0)).
\end{equation*} Moreover,
\begin{equation*}
\frac{\pa U}{\pa w}= -\frac{\frac{\pa \mathcal{A}}{\pa w}}{\frac{\pa \mathcal{A}}{\pa\tilde{w}}}, \hspace{1cm}\frac{\pa U}{\pa t_0} =\frac{1}{\frac{\pa \mathcal{A}}{\pa\tilde{w}}}.
\end{equation*}Define the function $V(w, t_0)$ by
\begin{equation*}
V(w, t_0)=\tilde{\mathcal{A}}(w,U(w, t_0)).
\end{equation*}The relations \eqref{eq6_5_14} are then equal to \eqref{eq6_5_11}. On the other hand,
\begin{equation*}\begin{split}
\frac{\pa V}{\pa w} =&\frac{\pa\tilde{\mathcal{A}}}{\pa w} +\frac{\pa \tilde{\mathcal{A}}}{\pa \tilde{w} }\frac{\pa U}{\pa w}=\frac{\pa\tilde{\mathcal{A}}}{\pa w} -\frac{\pa \tilde{\mathcal{A}}}{\pa \tilde{w} }\frac{\frac{\pa \mathcal{A}}{\pa w}}{\frac{\pa \mathcal{A}}{\pa\tilde{w}}},\\
\frac{\pa V}{\pa t_0}=&\frac{\pa \tilde{A}}{\pa \tilde{w}}\frac{\pa U}{\pa t_0}=\frac{\frac{\pa \tilde{A}}{\pa \tilde{w}}}{\frac{\pa \mathcal{A}}{\pa\tilde{w}}}.
\end{split}\end{equation*}Therefore,
\begin{equation*}
\left\{ U, V\right\}= w\frac{\pa U}{\pa w}\frac{\pa V}{\pa t_0} -w\frac{\pa U}{\pa t_0}\frac{\pa V}{\pa w}= -w\frac{\frac{\pa \tilde{\mathcal{A}}}{\pa w}}{\frac{\pa\mathcal{A}}{\pa \tilde{w}}}=\tilde{w}=U(w, t_0).
\end{equation*}In other words, $(\mathcal{L}, \tilde{\mathcal{L}})$ is a solution of the dispersionless Toda hierarchy with Orlov-Schulman functions $\mathcal{M}(w)$ and $\tilde{\mathcal{M}}(w)$.

Summarizing, we have shown the following.
\begin{proposition}
For  $(\mathcal{L}, \tilde{\mathcal{L}})$ to be a solution of the dispersionless Toda hierarchy with Orlov-Schulman functions $\mathcal{M}(w)$ and $\tilde{\mathcal{M}}(w)$, either one of the following cases holds:
\\
\textbf{Case I} \;\; $\{\mathcal{L},\tilde{\mathcal{L}}\}=0$. In this case, there exists two   functions $U$ and $U_1$ of $z$ such that $U'(z)\neq 0$ and
\begin{equation*}
\tilde{\mathcal{L}}=U(\mathcal{L}), \hspace{1cm} \tilde{\mathcal{M}}= \frac{U(\mathcal{L})}{\mathcal{L}U'(\mathcal{L})}\mathcal{M}+U_1(\mathcal{L}).
\end{equation*}
\\
\textbf{Case II} \;\; $\{\mathcal{L},\tilde{\mathcal{L}}\}\neq 0$. In this case, there exists a function $\mathcal{H}(z_1, z_2)$ such that
$\pa_{z_1}\pa_{z_2} \mathcal{H}(z_1, z_2)\neq 0$, and
\begin{equation}\label{eq6_8_1}
\mathcal{M}= \mathcal{L}\pa_{z_1}\mathcal{H}\left(\mathcal{L}, \tilde{\mathcal{L}}\right), \hspace{1cm}\tilde{\mathcal{M}}= -\tilde{\mathcal{L}}\pa_{z_2}\mathcal{H}\left(\mathcal{L}, \tilde{\mathcal{L}}\right).
\end{equation}In this case, the string equation is
\begin{equation*}
\left\{\mathcal{L}, \tilde{\mathcal{L}}\right\}  =\frac{1}{\pa_{\mathcal{L}}\pa_{\tilde{\mathcal{L}}}\mathcal{H}(\mathcal{L}, \tilde{\mathcal{L}})}.
\end{equation*}
\end{proposition}

Conversely, we have
\begin{proposition}
If $(\mathcal{L}, \tilde{\mathcal{L}})$ are functions of the form  \eqref{eq6_5_6},   $\mathcal{M}, \tilde{\mathcal{M}}$ are functions of the form \eqref{eq6_5_7}, and if $U$ and $U_1$ are two functions of $z$ so that $U'(z)\neq 0$ and
\begin{equation*}
\tilde{\mathcal{L}}=U(\mathcal{L}), \hspace{1cm} \tilde{\mathcal{M}}= \frac{U(\mathcal{L})}{\mathcal{L}U'(\mathcal{L})}\mathcal{M}+U_1(\mathcal{L}),
\end{equation*} then $(\mathcal{L}, \tilde{\mathcal{L}})$ is a solution of the dispersionless Toda hierarchy with Orlov-Schulman functions $\mathcal{M}(w)$ and $\tilde{\mathcal{M}}(w)$.
\end{proposition}

\begin{proposition}
If $(\mathcal{L}, \tilde{\mathcal{L}})$ are functions of the form  \eqref{eq6_5_6},   $\mathcal{M}, \tilde{\mathcal{M}}$ are functions of the form \eqref{eq6_5_7}, and if $\mathcal{H}(z_1, z_2)$ is a function of $z_1$ and $z_2$ such that
$\pa_{z_1}\pa_{z_2} \mathcal{H}(z_1, z_2)\neq 0$, and
\begin{equation*}
\mathcal{M}= \mathcal{L}\pa_{z_1}\mathcal{H}\left(\mathcal{L}, \tilde{\mathcal{L}}\right), \hspace{1cm}\tilde{\mathcal{M}}= -\tilde{\mathcal{L}}\pa_{z_2}\mathcal{H}\left(\mathcal{L}, \tilde{\mathcal{L}}\right),
\end{equation*}then $(\mathcal{L}, \tilde{\mathcal{L}})$ is a solution of the dispersionless Toda hierarchy with Orlov-Schulman functions $\mathcal{M}(w)$ and $\tilde{\mathcal{M}}(w)$.

\end{proposition}

From the results above, we see that we can use the string equation to classify the dispersionless Toda hierarchy into degenerate and non-degenerate cases. $(\mL, \tilde{\mL})$ is a degenerate solution if and only if $$\left\{\mL, \tilde{\mL}\right\}=0.$$However, the string equation does not determine the solution uniquely. In the degenerate case, we see that the solutions are determined by two  functions $U$ and $U_1$. In the non-degenerate case, i.e., when $$\{\mL, \tilde{\mL}\}=\frac{1}{\pa_{\mathcal{L}}\pa_{\tilde{\mathcal{L}}}\mathcal{H}(\mathcal{L}, \tilde{\mathcal{L}})}\neq 0,$$ the solution is also only determined up to two arbitrary functions. More precisely, for any two functions $\mathcal{H}_1(z_1)$ and $ \mathcal{H}_2(z_2)$ of $z_1$ and $z_2$  respectively, the functions $\mathcal{H}(z_1, z_2)$ and $\mathcal{H}(z_1, z_2)+\mathcal{H}_1(z_1)+ \mathcal{H}_2(z_2)$ give rise to the same string equation. In the following section, we are mainly going to discuss the non-degenerate solutions of the dispersionless Toda hierarchy. We are going to show that the roles of the two   auxiliary functions $\mathcal{H}_1(z_1)$ and $\mathcal{H}_2(z_2)$ are just shifting the origin of the time coordinates $t_n, n\in\Z$.

\section{Dispersionless Toda flows on space of pairs of conformal mappings}\label{sec4}

Let $\Del$ and $\Del^*$ be respectively the unit disc  and its exterior. As in \cite{1}, we introduce the following spaces of conformal mappings.

\begin{equation*}\begin{split}
\mathfrak{S}_{\infty}=&\Bigl\{ g: \mathbb{D}^* \rightarrow \C \;\text{univalent}
\;\bigr\vert\;
g(w)=b w+b_0 + b_1w^{-1}+\ldots; b\neq0; \\
&\hspace{0.5cm} 0\notin g(\Del^*);\;\;g \;\text{is extendable to a $C^1$   homeomorphism
of $\hat{\C}$}.\Bigr\},\\
\mathfrak{S}_0=&\Bigl\{ f: \mathbb{D} \rightarrow \C \;\text{univalent}
\;\bigr\vert\;
f(w)=a_1 w+a_2w^2+ \ldots; a_1\neq0; \\
&\hspace{0.5cm} \infty \notin f(\Del);\;\;f \;\text{is extendable to a $C^1$   homeomorphism
of $\hat{\C}$}.\Bigr\},\\
\mathfrak{D}=&\Bigl\{(g,f)\; \bigr\vert\; g\in \mathfrak{S}_{\infty},  f\in \mathfrak{S}_0;\;
f'(0)g'(\infty) = a_1b=1 .\Bigr\}.\end{split}
\end{equation*}Let $\Omega_1^-= g(\mathbb{D}^*)$ and $\Omega_1^+$ its exterior,
$\Omega_2^+=f(\mathbb{D})$ and $\Omega_2^-$ its exterior. $\mathcal{C}_1$ and $\mathcal{C}_2$ denotes the $C^1$ curves $\mathcal{C}_1=g(S^1)$ and $\mathcal{C}_2=f(S^1)$ respectively\footnote{The notations used here is a little bit different from those used in \cite{1}.}.

The main objective of this paper is to interpret the dispersionless Toda flows as integrable structure on the space $\mathfrak{D}$ of pairs of conformal mappings, by identifying $\mathcal{L}(w)$ with $g(w)$ and $\tilde{\mL}(w)$ with $f(w)$. As discussed in the previous section, the solutions of the dispersionless Toda hierarchy can be classified into degenerate and non-degenerate solutions. For the degenerate solutions, $\tilde{\mL}$ can be expressed as a nontrivial function $U$ of  $\mL$, which is independent of the time variables. This implies that the dispersionless Toda flows are restricted to the subspace of $\mathfrak{D}$ defined by $f(w)=U(g(w))$, and the time variables $t_n, n\in \Z$, will not be independent. Therefore,   we will not study the degenerate flows in this paper. We focus on the non-degenerate flows where the time variables are locally coordinates of the space $\mathfrak{D}$.

Although the results in the previous section shows that any function $\mathcal{H}(z_1,z_2)$ with $\pa_{z_1}\pa_{z_2}\mathcal{H}(z_1, z_2)\neq0$ gives rise to a solution of the dispersionless Toda hierarchy, this is not the end of the story.  In this section, we are going to show that given $\mathcal{H}(z_1,z_2)$ with $\pa_{z_1}\pa_{z_2}\mathcal{H}(z_1, z_2)\neq0$, how the time variables $t_n, n\in\Z$, are defined in terms of $g(w)$ and $f(w)$. We will prove that these time coordinates   can play the role of local coordinates on the space $\mathfrak{D}$. We are also going to define the phi function $\phi$ and the tau function $\tau$ in terms of $\mathcal{H}(z_1, z_2)$. To be more concrete, one can assume that
$$\mathcal{H}(z_1,z_2)= z_1^{\mu}z_2^{-\nu}, \hspace{1cm}\mu,\nu\in \mathbb{Z}\setminus\{0\},$$ or a linear combination of these functions. However, the results of this section does not depend on the specific form of the function $\mathcal{H}(z_1,z_2)$. Nevertheless, we assume throughout that $\mathcal{H}(z_1,z_2)$ is an analytic function of $z_1$ and $z_2$ on $\mathbb{C}\setminus\{0\}$. The case considered in \cite{1} is the special case where $\mathcal{H}(z_1,z_2)=z_1z_2^{-1}$.

We begin with the definitions of $t_n, n\in \Z$, and $v_n, n\in \Z$, in terms of $\mathcal{H}(z_1, z_2)$, and $g(w)$ and $f(w)$. From \eqref{eq6_5_7} and \eqref{eq6_8_1}, we find that $t_n, n\in \Z$, and $v_n, n\in \Z$, should be defined as: For $n\geq 1$,
\begin{equation}\label{eq6_8_2}
\begin{split}
  t_n=&\frac{1}{2\pi i n}\oint_{S^1}\pa_{z_1}\mathcal{H}(g(w), f(w))g(w)^{-n} dg(w)=\frac{1}{2\pi i n}\oint_{\mathcal{C}_1}\pa_{z_1}\mathcal{H}(z, f\circ g^{-1}(z))z^{-n} dz,  \\
t_{-n}=&\frac{1}{2\pi i n }\oint_{S^1} \pa_{z_2}\mathcal{H}(g(w), f(w))f(w)^{n}df(w)=\frac{1}{2\pi i n }\oint_{\mathcal{C}_2} \pa_{z_2}\mathcal{H}(g\circ f^{-1}(z), z)z^{n}dz,\\
v_n=&\frac{1}{2\pi i }\oint_{S^1}\pa_{z_1}\mathcal{H}(g(w), f(w))g(w)^{n} dg(w)=\frac{1}{2\pi i }\oint_{\mathcal{C}_1}\pa_{z_1}\mathcal{H}(z, f\circ g^{-1}(z))z^{n} dz,\\
v_{-n}=&\frac{1}{2\pi i  }\oint_{S^1} \pa_{z_2}\mathcal{H}(g(w), f(w))f(w)^{-n}df(w)=\frac{1}{2\pi i }\oint_{\mathcal{C}_2} \pa_{z_2}\mathcal{H}(g\circ f^{-1}(z), z)z^{-n}dz.\end{split}
\end{equation} The function $t_0$ is defined as\begin{equation}\begin{split}\label{eq6_8_3}t_0=& \frac{1}{2\pi i }\oint_{S^1}\pa_{z_1}\mathcal{H}(g(w), f(w)) dg(w)=\frac{1}{2\pi i }\oint_{\mathcal{C}_1}\pa_{z_1}\mathcal{H}(z, f\circ g^{-1}(z)) dz\\=&-\frac{1}{2\pi i  }\oint_{S^1} \pa_{z_2}\mathcal{H}(g(w), f(w)) df(w)=-\frac{1}{2\pi i  }\oint_{\mathcal{C}_2} \pa_{z_2}\mathcal{H}(g\circ f^{-1}(z), z) dz.
\end{split}
\end{equation}The function $v_0:=\phi$ will be defined later. Since $t_n,n\in \Z$, only depends explicitly  on  $g(w)$ and $f(w)$, but not their complex conjugates. Therefore, they are holomorphic functions on $\mathfrak{D}$.

In the following, we are going to show that the set of variables $t_n, n\in \Z$, are local coordinates on $\mathfrak{D}$, by showing that there are independent vector fields $\pa_n, n\in\Z$, on $\mathfrak{D}$ such that $\pa_n t_m=\delta_{n,m}$. For this purpose, we define the functions $S_+(z)$, $S_-(z)$, $\tilde{S}_+(z)$ and $\tilde{S}_-(z)$ by
\begin{equation}\label{eq6_9_1}\begin{split}
S_{\pm}(z) = &\frac{1}{2\pi i} \oint_{\mathcal{C}_1} \frac{\pa_{z_1}\mathcal{H}(\zeta, f\circ g^{-1}(\zeta))}{\zeta -z}d\zeta, \hspace{1cm} z\in \Omega_1^{\pm},\\
\tilde{S}_{\pm}(z) = &-\frac{1}{2\pi i} \oint_{\mathcal{C}_2} \frac{\pa_{z_2}\mathcal{H}( g\circ f^{-1}(\zeta),\zeta)}{\zeta -z}d\zeta, \hspace{1cm} z\in \Omega_2^{\pm}.
\end{split}
\end{equation}It is easy to see that in a neighbourhood of $z=0$, we have
\begin{equation}\label{eq6_9_2}
S_+(z)=\sum_{n=1}^{\infty}nt_n z^{n-1}, \hspace{1cm} \tilde{S}_+(z) =-\sum_{n=1}^{\infty} v_{-n}z^{n-1}.
\end{equation}In a neighbourhood of $z=\infty$,
\begin{equation}\label{eq6_9_3}
S_-(z)=-\frac{t_0}{z} -\sum_{n=1}^{\infty}v_nz^{-n-1}, \hspace{1cm}\tilde{S}_-(z)=-\frac{t_0}{z}+\sum_{n=1}^{\infty} nt_{-n}z^{-n-1}.
\end{equation}

Let $G$ and $F$ be the inverse functions of $g$ and $f$ respectively. Denote by $b_{m,n}, m,n\in\Z$, and $P_n(w), n\in \Z$, the  generalized Grunsky coefficients and Faber polynomials of $(G,F)$.
\begin{proposition}
There are independent vector fields $\pa_n, n\in \Z$, on $\mathfrak{D}$ such that $\pa_n t_m=\delta_{n,m}$.
\end{proposition}
\begin{proof}
We begin by constructing the vector fields $\pa_n, n\in \Z$. First, define the functions $\mathrm{u}_n(w), w\in S^1, n\in \Z$, by
\begin{equation}\label{eq6_8_7}
\begin{split}
\mathrm{u}_n(w) = -\frac{P_n'(w)}{f'(w)g'(w) \pa_{z_1}\pa_{z_2}\mathcal{H}(g(w),f(w))}=\sum_{m=-\infty}^{\infty}\mathrm{u}_{n;m}w^{m+1}.
\end{split}
\end{equation}Given a vector field $\pa$ on $\mathfrak{D}$, notice that
\begin{equation*}
\begin{split}
\frac{\pa g(w)}{g'(w)}=&( \pa \log b) w +\,\text{lower power terms in $w$},\\
\frac{\pa f(w)}{f'(w)} =& (\pa \log a_1) w+ \,\text{higher power terms in $w$}.
\end{split}
\end{equation*}Since $a_1b=1$, this implies that $\pa\log a_1=-\pa \log b$. Therefore, if we define vector fields $\pa_n$ on $\mathfrak{D}$ by
\begin{equation}\label{eq6_8_8}
\begin{split}
\pa_n g(w)= g'(w)\left(\frac{1}{2} \mathrm{u}_{n;0}w +\sum_{m=1}^{\infty} \mathrm{u}_{n;-m}w^{-m+1}\right),\\
\pa_n f(w) = f'(w) \left( -\frac{1}{2} \mathrm{u}_{n;0}w-\sum_{m=1}^{\infty} \mathrm{u}_{n;m}w^{m+1}\right),
\end{split}
\end{equation}then
\begin{equation}\label{eq6_9_8}
\frac{\pa_ng(w)}{g'(w)}-\frac{\pa_n f(w)}{f'(w)}=\mathrm{u}_n(w)=-\frac{P_n'(w)}{f'(w)g'(w) \pa_{z_1}\pa_{z_2}\mathcal{H}(g(w),f(w))}.
\end{equation}Since the functions $P_n'(w), n\in \Z,$ are independent, the functions $\mathrm{u}_n(w), n\in \Z$, are   independent. Therefore the vector fields $\pa_n, n\in\Z$, are also independent. In fact, we can say more. Since $P_0'(w) =1/w$, and for $n\geq 1$, $P_n'(w)\sim w^{n-1}+\,\text{lower positive power terms in $w$}$, $P_{-n}'(w)\sim w^{-n-1}+\,\text{lower negative power terms in $w$}$, we can conclude that the vector fields $\pa_n$ span the tangent space of $\mathfrak{D}$.

Now   for $z\in \mathcal{C}_1$,
\begin{equation}\label{eq6_8_5}
\begin{split}
  \pa_n \left( \pa_{z_1}\mathcal{H}\left(z, f\circ g^{-1}(z)\right)\right)
=&\pa_{z_1}\pa_{ z_2}\mathcal{H}\left(z, f\circ g^{-1}(z)\right) \left(\pa_nf -\frac{f'}{g'}\pa_ng\right)\circ g^{-1}(z)\\
=& P_n'(g^{-1}(z))(g^{-1})'(z).
\end{split}
\end{equation}
Using \eqref{eq6_9_1}, \eqref{eq6_9_2}, \eqref{eq6_9_3} together with \eqref{eq6_8_5} and \eqref{eq6_8_4}, we find that in a neighbourhood of $z=0$,
\begin{equation}\label{eq6_9_4}\begin{split}
\sum_{m=1}^{\infty}m(\pa_n t_m) z^{m-1}=\frac{1}{2\pi i}\oint_{\mathcal{C}_1} \frac{P_n'(g^{-1}(\zeta))(g^{-1})'(\zeta)}{\zeta-z}d\zeta =\begin{cases} nz^{n-1}, \hspace{0.5cm}&\text{if}\;\;n\geq 1,\\
0,&\text{if}\;\;n\leq 0.\end{cases}
\end{split}
\end{equation}In a neighbourhood of $z=\infty$,
\begin{equation}\label{eq6_9_5}
\begin{split}
\frac{\pa_n t_0}{z}+\sum_{m=1}^{\infty}(\pa_n v_m) z^{-m-1}=&\frac{-1}{2\pi i}\oint_{\mathcal{C}_1} \frac{P_n'(g^{-1}(\zeta))(g^{-1})'(\zeta)}{\zeta-z}d\zeta\\ =&\begin{cases} z^{-1} +\sum_{m=1}^{\infty}m
b_{0,m} z^{-m-1},\hspace{0.5cm}& \text{if}\;\; n=0,\\
  - n\sum_{m=1}^{\infty}m b_{nm}, &\text{if}\;\;n\geq 1,  \\
   - n\sum_{m=1}^{\infty}m
b_{-n,m} z^{-m-1}, &\text{if}\;\;n\leq -1.\end{cases}
\end{split}
\end{equation}

Similarly, for $z\in \mathcal{C}_2$, \begin{equation}\label{eq6_8_6}
\begin{split}
 - \pa_n \left( \pa_{z_2}\mathcal{H}\left(g\circ f^{-1}(z),z\right)\right)=&
-\pa_{z_1}\pa_{ z_2}\mathcal{H}\left(  g\circ f^{-1}(z), z\right)\left( \pa_n g -\frac{g'}{f'}\pa_n f\right)\circ f^{-1}(z)\\
=&P_n'(f^{-1}(z))(f^{-1})'(z).
\end{split}
\end{equation}
This, together with \eqref{eq6_9_1}, \eqref{eq6_9_2}, \eqref{eq6_9_3} and \eqref{eq6_8_4} imply that in a neighbourhood of $z=0$,
\begin{equation}\label{eq6_9_6}
\begin{split}
\sum_{m=1}^{\infty}(\pa_nv_{-m}) z^{m-1} = \begin{cases}
 \sum_{m=1}^{\infty}
m b_{0,-m} z^{m-1},\hspace{0.5cm}& \text{if}\;\; n=0,\\
 -n\sum_{m=1}^{\infty}m
b_{n, -m} z^{m-1}, &\text{if}\;\;n\geq 1,  \\
  -n\sum_{m=1}^{\infty}m b_{-n,-m} z^{m-1},&\text{if}\;\;n\leq -1.
\end{cases}
\end{split}
\end{equation}In a neighbourhood of $z=\infty$,
\begin{equation}\label{eq6_9_7}
\begin{split}
-\frac{\pa_n t_0}{z} +\sum_{m=1}^{\infty} m (\pa_n t_{-m}) z^{-m-1} =\begin{cases} -z^{-1}, \hspace{0.5cm}& \text{if}\;\; n=0,\\
0, &\text{if}\;\;n\geq 1,  \\
nz^{-n-1},&\text{if}\;\;n\leq -1.
\end{cases}
\end{split}
\end{equation}

Compare the coefficients on both sides of \eqref{eq6_9_4}, \eqref{eq6_9_5} and \eqref{eq6_9_7}, we find that
$$\pa_n t_m =\delta_{n,m}\hspace{1cm}\text{for all}\;\; n,m\in \Z,$$
which is the assertion of the proposition. As a result, $t_n, n\in \Z$, can play the role of local coordinates on $\mathfrak{D}$. The vector fields $\pa_n, n\in\Z$, constructed above are exactly the partial derivatives $\pa/\pa t_n$. From \eqref{eq6_9_5} and \eqref{eq6_9_6},  we also obtain
\begin{equation}\label{eq6_9_10}
\begin{split}
\frac{\pa v_m}{\pa t_n} = \begin{cases} -|mn|b_{m,n}, \hspace{0.5cm}&\text{if}\;\;n\neq 0,\\
|m|b_{m,0}, &\text{if}\;\; n=0,\end{cases}\hspace{1cm}m\neq 0.
\end{split}
\end{equation}
\end{proof} From \eqref{eq6_8_7} and \eqref{eq6_8_8}, we observe that the partial derivatives
\begin{equation}\label{eq6_8_9}
\frac{\pa g (w)}{\pa t_n}, \hspace{1cm}\frac{\pa f(w)}{\pa t_n}
\end{equation}depend on $\mathcal{H}(z_1, z_2)$ only through $\pa_{z_1}\pa_{z_2}\mathcal{H}(z_1, z_2)$. Therefore, these partial derivatives are not changed if we replace $\mathcal{H}(z_1, z_2)$ by $\mathcal{H}(z_1, z_2)+\mathcal{H}_1(z_1)+\mathcal{H}_2(z_2)$. Since the partial derivatives \eqref{eq6_8_9} determine $g(w)$ and $f(w)$ up to the initial condition, we can conclude that the roles of the two auxiliary functions $\mathcal{H}_1(z_1)$ and $\mathcal{H}_2(z_2)$ are to change the origin of the coordinates $t_n, n\in \Z$. In fact, this can also be observed in a different way. From the definitions \eqref{eq6_8_2} and \eqref{eq6_8_3}, we find that if we replace $\mathcal{H}(z_1, z_2)$ by $\mathcal{H}'(z_1, z_2)=\mathcal{H}(z_1, z_2)+\mathcal{H}_1(z_1)+\mathcal{H}_2(z_2)$, then the time variables $t_n, n\in \Z$, are changed to $t_n', n\in \Z$, where $$t_n'=t_n +c_n, \hspace{1cm} c_n = \begin{cases} \frac{1}{2\pi i n}\oint_{\infty} \pa_z\mathcal{H}_1(z)z^{-n}dz, \hspace{0.5cm}&\text{if}\;\;n\geq 1,\\
0, &\text{if}\; n=0,\\
\frac{1}{2\pi in}\oint_0 \pa_z\mathcal{H}_2(z) z^{n} dz,\hspace{0.5cm}&\text{if}\;\;n\leq -1.
\end{cases}$$In fact, the functions $v_n,n\in\Z\setminus\{0\}$ are also changed to $v_n'$, where
$$v_n'=v_n +d_n, \hspace{1cm} d_n = \begin{cases} \frac{1}{2\pi i }\oint_{\infty} \pa_z\mathcal{H}_1(z)z^{n}dz, \hspace{0.5cm}&\text{if}\;\;n\geq 1,\\
\frac{1}{2\pi i}\oint_0 \pa_z\mathcal{H}_2(z) z^{-n} dz,\hspace{0.5cm}&\text{if}\;\;n\leq -1.
\end{cases}$$Since $d_n,n\in \Z\setminus\{0\}$ are independent of $\boldsymbol{t}$, we find that replacing $\mathcal{H}(z_1, z_2)$ by $\mathcal{H}'(z_1, z_2)=\mathcal{H}(z_1, z_2)+\mathcal{H}_1(z_1)+\mathcal{H}_2(z_2)$ does not change the relations \eqref{eq6_9_10} which characterize the functions $v_n$ up to constants.

 We have used the forms of the Orlov-Schulman functions and the Riemann-Hilbert data to help us define the   variables $v_n, n\neq 0$, in terms of $f(w)$ and $g(w)$. Unfortunately, this does not lead to a definition of $v_0$, or equivalently, the phi function $\phi$, which should be defined so that
 \begin{equation}\label{eq6_8_10}
 \frac{\pa v_0}{\pa t_n} =\frac{\pa v_n}{\pa t_0}= \begin{cases} |n|b_{0,n},\hspace{0.5cm}&\text{if}\;\;n\neq 0,\\
 -2b_{0,0}, &\text{if}\;\; n=0.\end{cases}
 \end{equation}In other words, $v_0=\phi$ generates the coefficients of $b_{n,0}, n\in \mathbb{N}$, of $\log g^{-1}(z)$ and the coefficients $b_{-n,0}, n\in \mathbb{N}$, of $\log f^{-1}(z)$.

In the following, we define the function $v_0$ and show that it satisfies \eqref{eq6_8_10}.
\begin{proposition}
The function $v_0$ defined by
\begin{equation}\label{eq6_8_12}\begin{split}
 v_0=\frac{1}{2\pi i} \oint_{S^1}&\Biggl\{ \pa_{z_1}\mathcal{H}(g(w), f(w))g'(w) \log\frac{g(w)}{w} +\pa_{z_2}\mathcal{H}(g(w), f(w))f'(w) \log\frac{f(w)}{w}\\&  -\frac{\mathcal{H}(g(w), f(w))}{w}\Biggr\}dw\end{split}
\end{equation}satisfies
\begin{equation}\label{eq6_8_11}
 \frac{\pa v_0}{\pa t_n} = \begin{cases} |n|b_{0,n},\hspace{0.5cm}&\text{if}\;\;n\neq 0,\\
 -2b_{0,0}, &\text{if}\;\; n=0.\end{cases}
 \end{equation}
\end{proposition}
\begin{proof}
A straightforward computation gives
\begin{equation*}
\begin{split}
&\frac{\pa}{\pa t_n} \left\{\pa_{z_1}\mathcal{H}(g(w), f(w))g'(w) \log\frac{g(w)}{w} +\pa_{z_2}\mathcal{H}(g(w), f(w))f'(w) \log\frac{f(w)}{w}  -\frac{\mathcal{H}(g(w), f(w))}{w}\right\}\\=&
\frac{\pa}{\pa w}\left\{ \pa_{z_1}\mathcal{H}(g(w), f(w))\frac{\pa g(w)}{\pa t_n} \log\frac{g(w)}{w}+\pa_{z_2}\mathcal{H}(g(w), f(w))\frac{\pa f(w)}{\pa t_n}\log\frac{f(w)}{w}\right\}\\
&+f'(w)g'(w)\frac{\pa^2\mathcal{H}}{\pa z_1\pa z_2}(g(w), f(w))\left(\frac{\pa_n f(w)}{f'(w)}-\frac{\pa_n g(w)}{g'(w)}\right)\left(\log\frac{g(w)}{w}-\log\frac{f(w)}{w}\right).
\end{split}
\end{equation*}This together with the definition \eqref{eq6_8_12} of $v_0$ and \eqref{eq6_9_8}, give
\begin{equation}\label{eq6_9_9}
\begin{split}
\frac{\pa v_0}{\pa t_n}=\frac{1}{2\pi i}\oint_{S^1}P_n'(w)\left(\log\frac{g(w)}{w}-\log\frac{f(w)}{w}\right)dw.
\end{split}
\end{equation}For $n\geq 1$, \eqref{eq6_9_9} and \eqref{eq6_8_4} give
\begin{equation*}\begin{split}
\frac{\pa v_0}{\pa t_n}=&\frac{1}{2\pi i}\oint_{S^1} P_n'(w) \log\frac{g(w)}{w}dw= \frac{-1}{2\pi i}\oint_{\mathcal{C}_1} P_n'(g^{-1}(z))(g^{-1})'(z)\log \frac{g^{-1}(z)}{z}dz\\
=& \frac{-n}{2\pi i}\oint_{\mathcal{C}_1}\log \frac{g^{-1}(z)}{z} z^{n-1}dz=nb_{0,n},
\end{split}\end{equation*}
\begin{equation*}
\begin{split}
\frac{\pa v_0}{\pa t_{-n}}=&\frac{-1}{2\pi i}\oint_{S^1} P_{-n}'(w) \log\frac{f(w)}{w}dw=\frac{1}{2\pi i }\oint_{\mathcal{C}_2}P_{-n}'(f^{-1}(z))(f^{-1})'(z)\log\frac{f^{-1}(z)}{z}dz\\
=&\frac{-n}{2\pi i}\oint_{\mathcal{C}_2}\log\frac{f^{-1}(z)}{z}z^{-n-1}dz=nb_{0,-n}.
\end{split}
\end{equation*}For $n=0$, since $P_0'(w)=1/w$, we have
\begin{equation*}
\begin{split}
\frac{\pa v_0}{\pa t_0}=\frac{1}{2\pi i}\oint_{S^1}\left( \log\frac{g(w)}{w}-\log\frac{f(w)}{w}\right)\frac{1}{w}dw =\log b-\log a_1=2\log b =-2b_{0,0}.
\end{split}
\end{equation*}

\end{proof}
We would like to remark that if we replace $\mathcal{H}(z_1, z_2)$ by $\mathcal{H}'(z_1, z_2)=\mathcal{H}(z_1, z_2)+\mathcal{H}_1(z_1)+\mathcal{H}_2(z_2)$, then $v_0$ is changed to $v_0'$, where
\begin{equation*}
v_0'-v_0=\frac{1}{2\pi i}\oint_{\infty}\pa_z\mathcal{H}_1(z)\log z dz+\frac{1}{2\pi i}\oint_{0}\pa_2\mathcal{H}_2(z)\log z dz
\end{equation*}is independent of $\boldsymbol{t}$.

From \eqref{eq6_9_10}, \eqref{eq6_8_11} and the symmetry of the Grunsky coefficients, we find that
\begin{equation*}
\frac{\pa v_n}{\pa t_m}=\frac{\pa v_m}{\pa t_n} \hspace{1cm}\text{for all}\;\;n,m\in \Z.
\end{equation*}This gives the integrability condition for the tau function of the dispersionless Toda hierarchy, which is a function $\tau$ defined so that
\begin{equation}\label{eq6_10_10}
\begin{split}
\frac{\pa \log\tau}{\pa t_n}=v_n.
\end{split}
\end{equation} In the context of matrix models, the function $\log\tau$ is known as the free energy. Since in this paper, the variables $t_n, n\in\Z,$ are complex-valued, \eqref{eq6_10_10} does not determine $\log\tau$ uniquely. However, since the coefficients of $g(w)$ and $f(w)$ and the functions $v_n, n\in\Z$, depend holomorphically on $t_n$, it is natural to require $\log\tau$ to be a real-valued function so that
\begin{equation*}
\frac{\pa \log\tau}{\pa \bar{t}_n}=\bar{v}_n.
\end{equation*} 

Define the functions $\Phi(z), z\in \Omega_1^-$ and $\Psi(z), z\in \Omega_2^+$ so that  in a neighbourhood of $z=\infty$,
\begin{equation}\label{eq6_9_12}
\Phi(z)= \sum_{n=1}^{\infty}\frac{v_n}{n}z^{-n}.
\end{equation}In a neighbourhood of $z=0$,
\begin{equation}\label{eq6_9_13}
\Psi(z)=\sum_{n=1}^{\infty}\frac{v_{-n}}{n}z^n.
\end{equation}Notice that $\Phi'(z)= S_-(z)+t_0/z$ and $\Psi'(z)=-\tilde{S}_+(z)$.

Let $J_1(z_1,z_2)$ and $J_2(z_1, z_2)$ be two functions defined so that
\begin{equation}\label{eq6_10_3}
-\frac{\pa J_1}{\pa z_2}(z_1,z_2)=\frac{\pa J_2}{\pa z_1}(z_1,z_2)=\mathcal{H}(z_1,z_2)\frac{\pa^2 \mathcal{H}}{\pa z_1\pa z_2}(z_1,z_2).
\end{equation}We claim that the tau function of the dispersionless Toda hierarchy is given by
\begin{equation}
\tau =|\mathfrak{T}|^2,
\end{equation}where $\mathfrak{T}$ is a holomorphic function of $\boldsymbol{t}$ defined by
\begin{equation}\label{eq6_9_11}
\begin{split}
\log\mathfrak{T} =& \frac{t_0v_0}{2} +\frac{1}{4\pi i}\oint_{S^1}\Bigl\{ \pa_{z_1}\mathcal{H}(g(w), f(w)) g'(w) \Phi(g(w)) +\pa_{z_2}\mathcal{H}(g(w), f(w))f'(w)\Psi(f(w))\Bigr\}dw\\
&+ \frac{1}{8\pi i}\oint_{S^1}  \Bigl\{J_1(g(w),f(w))g'(w)+J_2(g(w),f(w))f'(w)\Bigr\}dw.
\end{split}
\end{equation}
Before proceed to the proof, we would like to comment that eq.  \eqref{eq6_10_3} only defines the function $J_1(z_1, z_2)$   up to a function of $z_1$ and the function  $J_2(z_1,z_2)$ up to a function of $z_2$. If we replace $J_1(z_1, z_2)$ and $J_2(z_1, z_2)$ by $J_1'(z_1,z_2)= J_1(z_1, z_2)+\tilde{J}_(z_1)$ and $J_2'(z_1,z_2)=J_2(z_1, z_2)+\tilde{J}_2(z_2)$ respectively, then
\begin{equation*}
\log \mathfrak{T}' =\log\mathfrak{T} + \frac{1}{8\pi i}\oint_{\infty}\tilde{J}_1(z)dz+ \frac{1}{8\pi i}\oint_{0} \tilde{J}_2(z)dz .
\end{equation*}Namely, $\log\tau'$ and $\log\tau$ only differ by a constant independent of $\boldsymbol{t}$. On the other hand, \eqref{eq6_10_3} implies that there exists a function $J_0(z_1, z_2)$ so that $$\frac{\pa}{\pa z_1}J_0(z_1, z_2)=-J_1(z_1, z_2), \hspace{1cm}\frac{\pa}{\pa z_2}J_0(z_1, z_2)=J_2(z_1, z_2).$$ Consequently,
\begin{equation*}
\begin{split}&\oint_{S^1}J_1(g(w),f(w))g'(w)dw=-\oint_{S^1}\pa_{z_1}J_0(g(w), f(w))g'(w)dw\\=&\oint_{S^1}\pa_{z_2}J_0(g(w),f(w))f'(w)dw=\oint_{S^1}J_2(g(w),f(w))f'(w)dw.
\end{split}\end{equation*}

Now we prove \eqref{eq6_10_10}.
\begin{proposition}
The function $\tau$ defined by \eqref{eq6_9_11} satisfies
\begin{equation}\label{eq6_10_6}
\begin{split}
\frac{\pa \log\tau}{\pa t_n}=v_n, \hspace{0.5cm} \frac{\pa \log\tau}{\pa \bar{t}_n}=\bar{v}_n, \hspace{1cm}\text{for all}\;\; n\in\Z.
\end{split}
\end{equation} Therefore it is the tau function of the dispersionless Toda hierarchy.
\end{proposition}
\begin{proof}
Since
$\log\tau =\log\mathfrak{T} +\overline{\log\mathfrak{T}}$ and $\log\mathfrak{T}$ is a holomorphic function of $\boldsymbol{t}$, it suffices to show that
\begin{equation*}
\frac{\pa\log\mathfrak{T}}{\pa t_n}=v_n \hspace{1cm}\text{for all}\;\; n\in\Z.
\end{equation*}
We write the function $\log\mathfrak{T}$ defined by \eqref{eq6_9_11} as the sum of three terms $\mathcal{Z}_1$, $\mathcal{Z}_2$ and $\mathcal{Z}_3$, where $\mathcal{Z}_1=t_0v_0/2$,
\begin{equation*}
\mathcal{Z}_2= \frac{1}{4\pi i}\oint_{S^1}\Bigl\{ \pa_{z_1}\mathcal{H}(g(w), f(w)) g'(w) \Phi(g(w)) +\pa_{z_2}\mathcal{H}(g(w), f(w))f'(w)\Psi(f(w))\Bigr\}dw,
\end{equation*}and
\begin{equation*}
\mathcal{Z}_3=\frac{1}{8\pi i}\oint_{S^1}  \Bigl\{J_1(g(w),f(w))g'(w)+J_2(g(w),f(w))f'(w)\Bigr\}dw.
\end{equation*}It follows immediately from \eqref{eq6_8_11} that
\begin{equation}\label{eq6_10_1}
\frac{\pa\mathcal{Z}_1}{\pa t_n}=\begin{cases} \frac{v_0}{2} -t_0b_{0,0},\hspace{0.5cm}&\text{if}\;\;n=0,\\
\frac{|n|}{2} t_0b_{0,n}, &\text{if}\;\; n\neq 0.
\end{cases}
\end{equation}For $\mathcal{Z}_2$, a straightforward computation gives
\begin{equation*}
\begin{split}
&\frac{\pa}{\pa t_n}\Bigl\{ \pa_{z_1}\mathcal{H}(g(w), f(w)) g'(w) \Phi(g(w)) +\pa_{z_2}\mathcal{H}(g(w), f(w))f'(w)\Psi(f(w))\Bigr\}\\
=&\frac{\pa}{\pa w}\left( \pa_{z_1}\mathcal{H}(g(w), f(w))\frac{\pa g(w)}{\pa t_n}\Phi(g(w))+\pa_{z_2}\mathcal{H}(g(w),f(w))\frac{\pa f(w)}{\pa t_n}\Psi(f(w))\right)\\
&+ g'(w)f'(w)\pa_{ z_1}\pa_{ z_2}\mathcal{H}(g(w),f(w))\left( \frac{\pa_n f(w)}{f'(w)}-\frac{\pa_n g(w)}{g'(w)}\right)\left(\Phi(g(w))-\Psi(f(w))\right)\\
&+\pa_{z_1}\mathcal{H}(g(w),f(w))g'(w)\pa_n\Phi(g(w))+\pa_{z_2}\mathcal{H}(g(w),f(w))f'(w)\pa_n\Psi(f(w)).
\end{split}
\end{equation*}Together with \eqref{eq6_9_8}, this implies that
\begin{equation*}
\begin{split}
&\frac{\pa\mathcal{Z}_2}{\pa t_n}=\frac{1}{4\pi i}\oint_{S^1} P_n'(w)\Bigl(\Phi(g(w))-\Psi(f(w))\Bigr)dw\\&+\frac{1}{4\pi i}\oint_{S^1}\Bigl(\pa_{z_1}\mathcal{H}(g(w),f(w))g'(w)\pa_n\Phi(g(w))+\pa_{z_2}\mathcal{H}(g(w),f(w))f'(w)\pa_n\Psi(f(w))\Bigr)dw.
\end{split}
\end{equation*}Now the definitions \eqref{eq6_9_12}, \eqref{eq6_9_13} imply that
\begin{equation}\label{eq6_10_2}
\frac{1}{4\pi i}\oint_{S^1} P_n'(w)\Bigl(\Phi(g(w))-\Psi(f(w))\Bigr)dw=\begin{cases}
0, \hspace{0.5cm}&\text{if}\;\; n=0,\\
v_n/2, &\text{if}\;\; n\neq 0.
\end{cases}
\end{equation}
On the other hand, the definitions \eqref{eq6_9_12}, \eqref{eq6_9_13} and the eqs. \eqref{eq6_9_10}, \eqref{eq6_5_2} and \eqref{eq6_5_3} imply that
\begin{equation*}
\begin{split}
\frac{\pa \Phi (z)}{\pa t_n} =\begin{cases}
 -\log\frac{g^{-1}(z)}{z} +b_{0,0}, \hspace{0.5cm} &\text{if}\;\; n=0,\\
- P_n(g^{-1}(z))+z^n, &\text{if}\;\;n\geq 1,\\
 -P_n(g^{-1}(z))+nb_{n,0}, &\text{if}\;\; n\leq -1,
\end{cases}
\end{split}
\end{equation*}
\begin{equation*}
\frac{\pa \Psi(z)}{\pa t_n}= \begin{cases}
-\log \frac{f^{-1}(z)}{z}-b_{0,0},  \hspace{0.5cm} &\text{if}\;\; n=0,\\
-P_n(f^{-1}(z))+nb_{n,0}, &\text{if}\;\;n\geq 1,\\
-P_n(f^{-1}(z))+z^n,&\text{if}\;\; n\leq -1.
\end{cases}
\end{equation*}Therefore,
\begin{equation*}
\begin{split}
&\frac{1}{4\pi i}\oint_{S^1}\Bigl(\pa_{z_1}\mathcal{H}(g(w),f(w))g'(w)\pa_n\Phi(g(w))+\pa_{z_2}\mathcal{H}(g(w),f(w))f'(w)\pa_n\Psi(f(w))\Bigr)dw\\
=&\begin{cases}
v_0/2+t_0b_{0,0}+\frac{1}{4\pi i}\oint_{S^1}\mathcal{H}(g(w),f(w))P_0'(w)dw, \hspace{1cm}&\text{if}\;\; n=0,\\
v_n/2-|n|t_0b_{n,0}/2+\frac{1}{4\pi i}\oint_{S^1} \mathcal{H}(g(w), f(w)) P_n'(w) dw, &\text{if}\;\;n\neq 0.
\end{cases}\end{split}
\end{equation*}Together with \eqref{eq6_10_1} and \eqref{eq6_10_2}, we find that
\begin{equation}\label{eq6_10_5}
\frac{\pa\mathcal{Z}_1}{\pa t_n}+\frac{\pa\mathcal{Z}_2}{\pa t_n}= v_n +\frac{1}{4\pi i}\oint_{S^1}  \mathcal{H}(g(w), f(w)) P_n'(w) dw.
\end{equation}Now we consider $\mathcal{Z}_3$. Using \eqref{eq6_10_3} and \eqref{eq6_9_8}, we have
\begin{equation*}
\begin{split}
&\frac{\pa}{\pa t_n}\Bigl\{J_1(g(w),f(w))g'(w)+J_2(g(w),f(w))f'(w)\Bigr\}\\=&
\frac{\pa}{\pa w}\left\{ J_1(g(w),f(w))\frac{\pa g(w)}{\pa t_n} + J_2(g(w),f(w))\frac{\pa f(w)}{\pa t_n}\right\}\\
&+2\mathcal{H}(g(w),f(w))f'(w)g'(w)\pa_{z_1}\pa_{z_2}\mathcal{H}(g(w),f(w))\left(\frac{\pa_ng(w)}{g'(w)}-\frac{\pa_n f(w)}{f'(w)}\right)\\=&
\frac{\pa}{\pa w}\left\{ J_1(g(w),f(w))\frac{\pa g(w)}{\pa t_n} + J_2(g(w),f(w))\frac{\pa f(w)}{\pa t_n}\right\}\\
&-2\mathcal{H}(g(w),f(w))P_n'(w).
\end{split}
\end{equation*}Therefore,
\begin{equation*}
\frac{\pa \mathcal{Z}_3}{\pa t_n}=-\frac{1}{4\pi i}\oint_{S^1}  \mathcal{H}(g(w), f(w)) P_n'(w) dw.
\end{equation*}Together with \eqref{eq6_10_5}, we conclude that
\begin{equation*}
\frac{\pa \log\tau}{\pa t_n}=\frac{\pa\log\mathfrak{T}}{\pa t_n}= v_n,
\end{equation*}which is the assertion of the proposition.
\end{proof}
Although it is not obvious from the definition, one can show that if the function $\mathcal{H}(z_1,z_2)$ is replaced by the function $\mathcal{H}(z_1, z_2)+\mathcal{H}_1(z_1)+\mathcal{H}_2(z_2)$, then the tau function $\tau$ is changed to $\tau'$, where $\log \tau'$ and $\log \tau$ differ by
$$\log\tau' -\log\tau =\sum_{n=-\infty}^{\infty}d_n t_n +\sum_{n=-\infty}^{\infty} \bar{d}_n \bar{t}_n+C,$$ which is a linear function of $\boldsymbol{t}$ and $ \bar{\boldsymbol{t}}$. On the other hand, using the definitions \eqref{eq6_9_12} and \eqref{eq6_9_13}, one can show that the function $\mathcal{Z}_2$ is given by
\begin{equation*}
\mathcal{Z}_2=\frac{1}{2}\left(\sum_{n=1} t_n v_n + \sum_{n=1}^{\infty} t_{-n} v_{-n}\right).
\end{equation*}Consequently, the holomorphic part of the tau function $\tau$ is
\begin{equation*}
 \mathfrak{T}=\exp\left(\frac{1}{2}\sum_{n=-\infty}^{\infty} t_n v_n + \frac{1}{8\pi i}\oint_{S^1}  \Bigl\{J_1(g(w),f(w))g'(w)+J_2(g(w),f(w))f'(w)\Bigr\}dw\right).
\end{equation*}

The equations \eqref{eq6_9_10}, \eqref{eq6_8_11} and \eqref{eq6_10_6} imply that the generalized Grunsky coefficients $b_{m,n}$ of the pair of univalent functions $(g^{-1}, f^{-1})$ can be generated by the tau function $\tau$:
\begin{equation}\label{eq6_10_9}\begin{split}
b_{m,n}= &
-\frac{1}{|mn|}\frac{\pa^2\log\tau}{\pa t_m\pa t_n}, \hspace{0.5cm}\text{if}\;\; mn\neq 0,\\
b_{m,n}= &\frac{1}{|m|}\frac{\pa^2\log\tau}{\pa t_m \pa t_0}, \hspace{1cm}\text{if}\;\;m\neq 0, n=0,\\
b_{m,n}= &\frac{1}{|n|}\frac{\pa^2\log\tau}{\pa t_0\pa t_n},\hspace{1cm}\text{if}\;\; m=0, n\neq 0,\\
b_{m,n}= &-\frac{1}{2}\frac{\pa^2\log\tau}{\pa t_0^2}, \hspace{0.8cm}\text{if}\;\;m=n=0.
\end{split} 
\end{equation} Therefore, \eqref{eq6_5_2} can be rewritten as
\begin{equation}\label{eq6_10_7}\begin{split}
\log\frac{g^{-1}(z)}{z} &= -\frac{1}{2}\frac{\pa^2\log\tau}{\pa t_0^2}-\sum_{m=1}^{\infty}
\frac{1}{m}\frac{\pa^2\log\tau}{\pa t_0\pa t_m} z^{-m}, \\\log\frac{f^{-1}(z)}{z} &=\frac{1}{2}\frac{\pa^2\log\tau}{\pa t_0^2}-\sum_{m=1}^{\infty}
\frac{1}{m}\frac{\pa^2\log\tau}{\pa t_0\pa t_{-m}} z^{m}.\end{split}
\end{equation}This shows that the coefficients of the conformal mappings can be expressed as second partial derivatives of the tau function. For $n\geq 1$, define
\begin{equation*}\begin{split}
\mathcal{B}_n(w)=& (g(w)^n)_{>0}+\frac{1}{2}(g(w)^n)_0=P_n(w)-\frac{n}{2}b_{n,0},\\ \mathcal{B}_{-n}(w)=&(f(w)^{-n})_{<0}+\frac{1}{2}(f(w)^{-n})_0=P_{-n}(w)+\frac{n}{2}b_{-n,0}.\end{split}
\end{equation*}
Then \eqref{eq6_5_3} and \eqref{eq6_10_9} imply that
\begin{equation}\label{eq6_10_8}\begin{split}
\mathcal{B}_n (g^{-1}(z))  =& z^{n} -\frac{1}{2}\frac{\pa^2\log\tau}{\pa t_n\pa t_0} -\sum_{m=1}^{\infty} \frac{1}{m}
\frac{\pa^2\log\tau}{\pa t_n\pa t_m}z^{-m},\\\mathcal{B}_n (f^{-1}(z)) =&
\frac{1}{2}\frac{\pa^2\log\tau}{\pa t_n\pa t_0} -\sum_{m=1}^{\infty}
\frac{1}{m}\frac{\pa^2\log\tau}{\pa t_n\pa t_{-m}} z^m,  \\
\mathcal{B}_{-n}(g^{-1}(z)) =& -\frac{1}{2}\frac{\pa^2\log\tau}{\pa t_{-n}\pa t_0} -\sum_{m=1}^{\infty}\frac{1}{m}
\frac{\pa^2\log\tau}{\pa t_{-n}\pa t_m} z^{-m},\\\mathcal{B}_{-n} (f^{-1}(z))=&
z^{-n} +\frac{1}{2}\frac{\pa^2\log\tau}{\pa t_{-n}\pa t_0}-\sum_{m=1}^{\infty} \frac{1}{m}\frac{\pa^2\log\tau}{\pa t_{-n}\pa t_{-m}} z^{m}.\end{split}
\end{equation}
From \eqref{eq6_10_7} and \eqref{eq6_10_8}, we find that for all $n\in \Z$,
\begin{equation*}
\begin{split}
\frac{\pa g^{-1}}{\pa t_n}(z) = &g^{-1}(z)\left( -\frac{1}{2}\frac{\pa^3\log\tau}{\pa t_n\pa t_0^2}-\sum_{m=1}^{\infty}
\frac{1}{m}\frac{\pa^2\log\tau}{\pa t_n\pa t_0\pa t_m} z^{-m}\right)=g^{-1}(z)\frac{\pa\mathcal{B}_n\circ g^{-1}}{\pa t_0}(z),\\
\frac{\pa f^{-1}}{\pa t_n}(z) =& f^{-1}(z)\left( \frac{1}{2}\frac{\pa^3\log\tau}{\pa t_n\pa t_0^2}-\sum_{m=1}^{\infty}
\frac{1}{m}\frac{\pa^2\log\tau}{\pa t_n\pa t_0\pa t_{-m}} z^{m}\right)=f^{-1}(z)\frac{\pa\mathcal{B}_n\circ f^{-1}}{\pa t_0}(z).
\end{split}
\end{equation*}Therefore, by chain rule, we find that for all $n\in \Z$,
\begin{equation*}\begin{split}
\frac{\pa g(w)}{\pa t_n}=&-\frac{\pa g(w)}{\pa w} \frac{\pa g^{-1}}{\pa t_n}\circ g(w)=-w\frac{\pa g(w)}{\pa w}\frac{\pa \mathcal{B}_n\circ g^{-1}}{\pa t_0}\circ g(w)  \\=&-w\frac{\pa g(w)}{\pa w}\left(\frac{\pa \mathcal{B}_n}{\pa t_0}\circ g^{-1} +\frac{\pa \mathcal{B}_n}{\pa w} \circ g^{-1}\frac{\pa g^{-1}}{\pa t_0}\right)\circ g(w)\\
=&w\frac{\pa\mathcal{B}_n(w)}{\pa w}\frac{\pa g(w)}{\pa t_0}-w\frac{\pa \mathcal{B}_n(w)}{\pa t_0}\frac{\pa g(w)}{\pa w}=\left\{\mathcal{B}_n(w), g(w)\right\}.\end{split}
\end{equation*}Similarly,
\begin{equation*}
\frac{\pa f(w)}{\pa t_n} = \left\{\mathcal{B}_n(w), f(w)\right\}.
\end{equation*}These are precisely the Lax equations of the dispersionless Toda hierarchy \eqref{eq6_5_4}. By setting $n=0$ in \eqref{eq6_9_8} and using the fact that $P_0'(w)=1/w$, we find that
\begin{equation}\label{eq6_16_1}
\Bigl\{ g(w), f(w)\Bigr\} = w g'(w)\frac{\pa f(w)}{\pa t_0} -wf'(w)\frac{\pa g(w)}{\pa t_0}=\frac{1}{\pa_{z_1}\pa_{z_2}\mathcal{H}(g(w),f(w))},
\end{equation}which is the string equation.

\section{Reduction to subspaces of $\mathfrak{D}$}
In this section, we consider the reduction of the dispersionless Toda hierarchy to some subspaces of $\mathfrak{D}$.
First we consider the subspace $\mathfrak{R}$ of $\mathfrak{D}$ consists of $(g,f)$ satisfying
\begin{equation*}
\overline{g(\bar{w})} =g(w), \hspace{1cm} \overline{f(\bar{w})} =f(w),
\end{equation*}or equivalently, $g$ and $f$ has real coefficients. Assume that $\mathcal{H}(z_1, z_2)$ is  real-valued when $z_1$ and $z_2$ are real.  It is easy to check from \eqref{eq6_8_2}, \eqref{eq6_8_3} and \eqref{eq6_8_12} that restricted to $\mathfrak{R}$, all the variables $t_n, v_n, n\in\Z$, are real-valued. Therefore,  the subspace $\mathfrak{R}$ can be defined by the condition $\bar{t}_n=t_n$ for all $n\in\Z$. In other words, if $\mathcal{H}(z_1, z_2)$ is real-valued when $z_1$ and $z_2$ are real,   the Toda flows $\pa/\pa t_n, n\in\Z,$ on $\mathfrak{D}$ naturally restrict to the subspace $\mathfrak{R}$ and give rise to solutions of the real-valued dispersionless Toda hierarchy. The tau function is given by $\mathfrak{T}$ defined in \eqref{eq6_9_11}.

Next we consider the subspace $\Sigma$ of $\mathfrak{D}$ consists of $(g,f)$ where
\begin{equation}\label{eq6_16_2}
f(w)=\frac{1}{\overline{g(1/\bar{w})}}.
\end{equation}In this case, if the function $\mathcal{H}(z_1, z_2)$ is defined such that
the function $$\mathcal{U}(z, \bar{z}) =\mathcal{H}\left(z, \bar{z}^{-1}\right)$$ is real-valued, then \eqref{eq6_8_2}, \eqref{eq6_8_3} and \eqref{eq6_8_12} show that restricted to $\Sigma$, if $n\geq 1$,
\begin{equation*}
\begin{split}
t_n =&\frac{1}{2\pi in}\oint_{S^1} \pa_{z_1}\mathcal{H}\left(g(w), \overline{g(w)}\,^{-1}\right)g(w)^{-n}dg(w) =\frac{1}{2\pi i n}\oint_{\mathcal{C}_1} \pa_z\mathcal{U}(z, \bar{z})z^{-n}dz,\\
t_{-n}=&\frac{1}{2\pi i n} \oint_{S^1}\pa_{z_2}\mathcal{H}\left(g(w), \overline{g(w)}\,^{-1}\right)\overline{g(w)}\,^{-n}d  \overline{g(w)}\,^{-1}=\frac{1}{2\pi i n}\oint_{ \mathcal{C}_1}\pa_{\bar{z}}\mathcal{U}(z,\bar{z})\bar{z}^{-n}d\bar{z},
\\
v_n =&\frac{1}{2\pi i}\oint_{\mathcal{C}_1}\pa_z\mathcal{U}(z,\bar{z}) z^n dz,\hspace{1cm}
v_{-n}=\frac{1}{2\pi i}\oint_{\mathcal{C}_1}\pa_{\bar{z}}\mathcal{U}(z, \bar{z})\bar{z}^nd\bar{z},
\end{split}
\end{equation*}and if $n=0$,
\begin{equation}\label{eq6_11_1}\begin{split}
t_0=&\frac{1}{2\pi i} \oint_{\mathcal{C}_1}\pa_z\mathcal{U}(z,\bar{z}) dz= \frac{-1}{2\pi i}\oint_{\mathcal{C}_1}\pa_{\bar{z}}\mathcal{U}(z,\bar{z})d\bar{z},\\
v_0=&\frac{1}{2\pi i}\oint_{\mathcal{C}_1}\Bigl( \pa_z\mathcal{U}(z,\bar{z})\log z dz -\pa_{\bar{z}}\mathcal{U}(z,\bar{z})\log\bar{z}d\bar{z}\Bigr).
\end{split}\end{equation}
It is easy to see that on the space $\Sigma$,
\begin{equation*}
t_{-n}=-\bar{t}_n, \;\;\;\;v_{-n}=-\bar{v}_n \hspace{0.5cm}\text{for}\;\;n\neq 0, \hspace{0.5cm}\text{and}\hspace{0.5cm}t_0=\bar{t}_0, \;\;\; v_0=\bar{v}_0.
\end{equation*}
Therefore the subspace $\Sigma$ is characterized by $t_{-n}=-\bar{t}_n$ for all $n\neq 0$ and $t_0=\bar{t}_0$. In particular, $t_0$ is real-valued. Let $$\frac{\pa}{\pa \mathrm{t}_n}=\frac{\pa}{\pa t_n} -\frac{\pa}{\pa \bar{t}_{-n}},\;\; n\geq 1, \hspace{1cm}\frac{\pa}{\pa\mathrm{t}_0}=\frac{\pa}{\pa t_{0}}+\frac{\pa}{\pa\bar{t}_0}.$$These are well-defined vector fields on the subspace $\Sigma$. Moreover, since $\mathfrak{T}$ and $v_n, n\in\Z$, are holomorphic functions of $t_n,n\in\Z$, on $\mathfrak{D}$, we find that restricted to $\Sigma$,  \begin{equation}\label{eq6_19_1}\frac{\pa\log\mathfrak{T}}{\pa \mathrm{t}_n}=\frac{\pa \log\mathfrak{T}}{\pa t_n} = v_n, \hspace{1cm}\frac{\pa v_m}{\pa \mathrm{t}_n}=\frac{\pa v_m}{\pa t_n}=\begin{cases}
-|mn|b_{m,n},\hspace{0.5cm}&\text{if}\;\; mn\neq 0,\\
|m|b_{m,0}, &\text{if}\;\;m\neq 0, n=0,\\
-2b_{0,0}, &\text{if}\;\; m=n=0.
\end{cases}\end{equation}Moreover, the restriction of the tau function $\mathfrak{T}$ \eqref{eq6_9_11} to $\Sigma$ is given by
\begin{equation}\label{eq6_11_3}
\begin{split}
\log\mathfrak{T}=&\frac{t_0v_0}{2}+\frac{1}{4\pi i}\oint_{\mathcal{C}_1} \left(\pa_zU(z, \bar{z}) \Phi(z) dz-\pa_{\bar{z}}U(z,\bar{z})\overline{\Phi(\bar{z})}d\bar{z}\right)\\&+\frac{1}{8\pi i}\oint_{\mathcal{C}_1} \left(\mathcal{V}_1(z, \bar{z})dz+\mathcal{V}_2(z,\bar{z})d\bar{z}\right),
\end{split}
\end{equation}where $\mathcal{V}_1(z,\bar{z})$ and $\mathcal{V}_2(z,\bar{z})$ are defined so that
\begin{equation*}
-\pa_{\bar{z}}\mathcal{V}_1(z,\bar{z})=\pa_z\mathcal{V}_2(z,\bar{z})=\mathcal{U}(z,\bar{z})\pa_{z}\pa_{\bar{z}}\mathcal{U}(z,\bar{z}).
\end{equation*}Since $\mathcal{U}(z,\bar{z})$ is real-valued, we can choose $\mathcal{V}_1(z,\bar{z})$ and $\mathcal{V}_2(z,\bar{z})$ such that they satisfy $\mathcal{V}_2(z,\bar{z}) =-\overline{ \mathcal{V}_1(z,\bar{z})}$. \eqref{eq6_11_3} then shows that $\log\mathfrak{T}$ is real-valued. 
\eqref{eq6_19_1} implies that
\begin{equation}\label{eq6_16_3}
\begin{split}\frac{\pa^2\log\mathfrak{T}}{\pa t_m \pa t_n}=&  -mn b_{m,n}, \hspace{0.5cm}  \frac{\pa^2\log\mathfrak{T}}{\pa \bar{t}_m \pa \bar{t}_n}=  -mn b_{-m,-n}, \hspace{0.5cm}
\frac{\pa^2\log\mathfrak{T}}{\pa t_m \pa \bar{t}_n}= mn b_{m,-n},  \\
\frac{\pa^2\log\mathfrak{T}}{\pa t_n\pa t_0} =&n b_{n,0}, \hspace{0.5cm}\frac{\pa^2\log\mathfrak{T}}{\pa \bar{t}_n\pa t_0}=-nb_{-n,0}, \hspace{0.5cm}\frac{\pa^2\log\mathfrak{T}}{\pa t_0^2} =2b_{0,0}.
\end{split}\end{equation}
We can then show as at the end of Section \ref{sec4}  that restricted to $\Sigma$,
\begin{equation*}
\frac{\pa g(w)}{\pa \mathrm{t}_n}=\left\{\mathcal{B}_n(w), g(w)\right\},\hspace{1cm}\frac{\pa  g(w)}{\pa \bar{\mathrm{t}}_n}= -\left\{\overline{\mathcal{B}_n(\bar{w})}, g(w)\right\},
\end{equation*}for all $n\geq 1$.
If we further assume that $\mathcal{U}(z, \bar{z})$ is regular at $z=0$, this is precisely the general conformal mapping problem considered by Zabrodin in \cite{31}. In this case, the functions $t_0$ and $v_0$ \eqref{eq6_11_1} can be rewritten as
\begin{equation*}
t_0=\frac{1}{\pi}\iint_{\Omega_1^+}\pa_{z}\pa_{\bar{z}}\mathcal{U}(z,\bar{z})dzd\bar{z},\hspace{1cm}v_0
=\frac{1}{\pi}\iint_{\Omega_1^+}\pa_{z}\pa_{\bar{z}}\mathcal{U}(z,\bar{z})\log|z|^2 dzd\bar{z}.
\end{equation*} Moreover, \eqref{eq6_11_3} can be further simplified to
\begin{equation*}\begin{split}
\log\mathfrak{T}=&\frac{t_0v_0}{2}+\frac{1}{2\pi}\iint_{\Omega_1^+} \pa_{z}\pa_{\bar{z}}\mathcal{U}(z,\bar{z})\left(\Phi(z)+\overline{\Phi(\bar{z})} \right)d^2z-\frac{1}{2\pi}\iint_{\Omega_1^+} \mathcal{U}(z,\bar{z})\pa_{z}\pa_{\bar{z}}\mathcal{U}(z,\bar{z})d^2z\\
=&\frac{1}{2}\left(t_0v_0+\sum_{n=1}^{\infty}t_n v_n +\sum_{n=1}^{\infty}\bar{t}_n\bar{v}_n\right)-\frac{1}{2\pi}\iint_{\Omega_1^+} \mathcal{U}(z,\bar{z})\pa_{z}\pa_{\bar{z}}\mathcal{U}(z,\bar{z})d^2z.
\end{split}\end{equation*}This agrees with the tau function derived in \cite{31} using electrostatic variational principle. The string equation \eqref{eq6_16_1} is equivalent to
\begin{equation*}
\Bigl\{ g(w), \overline{g(1/\bar{w})}\Bigr\}= \frac{1}{\pa_z\pa_{\bar{z}}\mathcal{U}\left(g(w), \overline{g(1/\bar{w})}\right)}.
\end{equation*}

The interest on the dispersionless Toda flows on the space $\Sigma$ is partly motivated by its intimate relations with the Dirichlet boundary value problem. This is first observed in \cite{7} and later discussed in detail in \cite{9}, for the special solution where $\mathcal{U}(z,\bar{z})=z\bar{z}$. For general $\mathcal{U}(z,\bar{z})$, such a relation also exists. Recall that for a simply connected domain $\Omega$ that contains the $\infty$, the Dirichlet boundary value problem seeks for a harmonic function $u(z)$ on $\Omega$ with prescribed boundary value $u_0(z)$ on the boundary $\mathcal{C}=\pa \Omega$ of $\Omega$. This problem has a unique solution which can be written in terms of the Dirichlet Green's function $G_{\Omega}(z_1, z_2)$:
\begin{equation*}
u(z)=-\frac{1}{2\pi}\oint_{\mathcal{C}}u_0(\zeta) \pa_n G_{\Omega}(z,\zeta)|d\zeta|,
\end{equation*}where $\pa_n$ is the normal derivative on $\mathcal{C}$, and the Dirichlet Green's function $G_{\Omega}(z_1, z_2)$ can be written in terms of the Riemann mapping $G=g^{-1}:\Omega\rightarrow \Del^*$ by
\begin{equation*}
G_{\Omega}(z_1,z_2)=\log\left|\frac{G(z_1)-G(z_2)}{G(z_1)\overline{G(z_2)}-1}\right|.
\end{equation*}
Notice that the relation \eqref{eq6_16_2} gives
\begin{equation*}
F(z)=\frac{1}{\overline{G\left(1/\bar{z}\right)}},
\end{equation*}where $F=f^{-1}$. Setting $F(z)=1/\overline{G(1/\bar{z})}$ in \eqref{eq6_5_1}, we find that
\begin{equation*}\begin{split}
\log \frac{\overline{G(z)}}{\bar{z}}= &\log\beta+\sum_{n=1}^{\infty} b_{0,-n}\bar{z}^{-n},\\
\log \frac{G(z)\overline{G(z_2)}-1}{z_1\bar{z}_2}=&2b_{0,0}+\sum_{n=1}^{\infty} b_{-n,0}\bar{z}_2^{-n}-\sum_{n=1}^{\infty} b_{n,0}z_1^{-n}-\sum_{m=1}^{\infty}\sum_{n=1}^{\infty}b_{m,-n}z_1^{-m}\bar{z}_2^{-n}.
\end{split}\end{equation*} Together with the other identities in \eqref{eq6_5_1} and \eqref{eq6_16_3}, we find that
\begin{equation}\label{eq6_16_4}
G_{\Omega}(z_1,z_2)=\log\left|\frac{G(z_1)-G(z_2)}{G(z_1)\overline{G(z_2)}-1}\right|=\log\left|\frac{1}{z_1}-\frac{1}{z_2}\right|+\frac{1}{2}D(z_1)D(z_2)\log\mathfrak{T},
\end{equation}where $D(z)$ is the operator
\begin{equation*}
D(z)=\frac{\pa}{\pa t_0}+\sum_{n=1}^{\infty}\frac{z^{-n}}{n}\frac{\pa}{\pa t_n}+\sum_{n=1}^{\infty}\frac{\bar{z}^{-n}}{n}\frac{\pa}{\pa \bar{t}_n}.
\end{equation*}In other words, the tau function $\mathfrak{T}$ can be used to generate the coefficients of the Dirichlet Green's function. Notice that \eqref{eq6_16_4} is independent of the function $\mathcal{U}(z,\bar{z})$ which specifies the solution of the dispersionless Toda hierarchy. In other words, it is an universal relation that has to be satisfied  for  any dispersionless Toda flows that is restricted to the space $\Sigma$. It can be regarded as the compact form of the dispersionless Hirota equations \cite{3,33} for the dispersionless Toda hierarchies on $\Sigma$.

\section{Special cases}
In this section, we consider the special case where $\mathcal{H}(z_1, z_2)=z_1^{\mu}z_2^{-\nu}$, $\mu,\nu$ are nonzero integers. The relation \eqref{eq6_8_1} is then equal to
\begin{equation}\label{eq6_16_5}
\mathcal{M}= \mu \mathcal{L}^{\mu}\tilde{\mathcal{L}}^{-\nu}, \hspace{1cm} \tilde{\mathcal{M}}=\nu \mathcal{L}^{\mu}\tilde{\mathcal{L}}^{-\nu},
\end{equation}or equivalently,
\begin{equation*}
\mathcal{L}^{\mu}=\frac{\tilde{\mathcal{M}}\tilde{\mathcal{L}}^{\nu}}{\nu}, \hspace{1cm} \tilde{\mathcal{L}}^{-\nu}=\frac{\mathcal{M}\mathcal{L}^{-\mu}}{\mu}.
\end{equation*}This Riemann-Hilbert problem was proposed by Takasaki in \cite{32}, and studied by Alonso and Medina in \cite{25}.

Using the definition of $\mathcal{M}$ and $\tilde{\mathcal{M}}$ \eqref{eq6_5_7}, and setting $\mathcal{L}(w)=g(w)$ and $\tilde{\mathcal{L}}(w)=f(w)$, \eqref{eq6_16_5} implies that
\begin{equation}\label{eq6_17_1}\begin{split}
\mu g(w)^{\mu}f(w)^{-\nu} =&\sum_{n=1}^{\infty} nt_n g(w)^{n}+t_0+\sum_{n=1}^{\infty} v_n g(w)^{-n},\\
\nu g(w)^{\mu}f(w)^{-\nu} =&-\sum_{n=1}^{\infty} nt_{-n} f(w)^{-n} +t_0-\sum_{n=1}^{\infty} v_{-n} f(w)^n.
\end{split}
\end{equation}Multiplying the first equation by $g'(w)/g(w)$ and the second equation by $f'(w)/f(w)$, subtracting  the two resulting equations and integrating with respect to $w$, we find that
\begin{equation}\label{eq6_17_2}\begin{split}
g(w)^{\mu}f(w)^{\nu}= &\sum_{n=1}^{\infty} t_n g(w)^n + t_0 \log \frac{g(w)}{w} -\sum_{n=1}^{\infty} \frac{v_{n}}{n}g(w)^{-n}\\&-\sum_{n=1}^{\infty} t_{-n}f(w)^{-n}-t_0\log \frac{f(w)}{w}+\sum_{n=1}^{\infty}\frac{v_{-n}}{n}f(w)^n.\end{split}
\end{equation}The functions $t_n, n\in\Z$, are given by
\begin{equation}\label{eq6_17_3}\begin{split}
t_0=&\frac{\mu}{2\pi i }\oint_{S^1} g(w)^{\mu-1}f(w)^{-\nu} dg(w)=\frac{\nu}{2\pi i}\oint_{S^1}g(w)^{\mu}f(w)^{-\nu-1}df(w), \\
t_n=&\frac{\mu}{2\pi i n}\oint_{S^1} g(w)^{\mu-n-1}f(w)^{-\nu} dg(w), \hspace{1cm} n\geq 1,\\
t_{-n}=&\frac{-\nu}{2\pi i n}\oint_{S^1} g(w)^{\mu} f(w)^{-\nu+n-1}df(w), \hspace{1cm} n\geq 1,\end{split}
\end{equation}and the functions $v_n, n\in\Z$, are given by
\begin{equation}\label{eq6_17_4}
\begin{split}
v_n =&\frac{\mu}{2\pi i}\oint_{S^1} g(w)^{\mu+n-1}f(w)^{-\nu} dg(w), \hspace{1cm} n\geq 1,\\
v_{-n}=&\frac{-\nu}{2\pi i}\oint_{S^1} g(w)^{\mu} f(w)^{-\nu-n-1}df(w), \hspace{1cm} n\geq 1,\\
v_0=&\frac{1}{2\pi i}\oint_{S^1}g(w)^{\mu}f(w)^{-\nu}\left(\mu\frac{g'(w)}{g(w)}\log\frac{g(w)}{w}-\nu\frac{f'(w)}{f(w)}\log\frac{f(w)}{w}-\frac{1}{w}\right)dw.
\end{split}
\end{equation}
The functions $J_1(z_1, z_2)$ and $J_2(z_1, z_2)$ in \eqref{eq6_10_3} can be chosen to be
\begin{equation*}
J_1(z_1,z_2)=-\frac{\mu}{2}z_1^{2\mu-1}z_2^{-2\nu},\hspace{1cm} J_2(z_1, z_2)=-\frac{\nu}{2}z_1^{2\mu}z_2^{-2\nu-1}.
\end{equation*}
Consequently the holomorphic part of $\log\tau$ is
\begin{equation*}
\log\mathfrak{T}=\frac{1}{2}\sum_{n=-\infty}^{\infty} t_n v_n -\frac{1}{16\pi i}\oint_{S^1} \left(\mu g(w)^{2\mu-1}f(w)^{-2\nu}g'(w)+\nu g(w)^{2\mu}f(w)^{-2\nu-1}f'(w)\right)dw.
\end{equation*}Using \eqref{eq6_17_1}, \eqref{eq6_17_3} and \eqref{eq6_17_4}, we find that
\begin{equation*}
\begin{split}
&\frac{1}{2\pi i} \oint_{S^1}\mu g(w)^{2\mu-1}f(w)^{-2\nu}g'(w)dw\\=&\frac{1}{2\pi i} \oint_{S^1}\left(\sum_{n=1}^{\infty} nt_n g(w)^{n}+t_0+\sum_{n=1}^{\infty} v_n g(w)^{-n}\right)g(w)^{\mu-1}f(w)^{-\nu}dg(w)\\
=&\frac{1}{\mu} \left(2 \sum_{n=1}^{\infty} nt_nv_n +t_0^2\right),
\end{split}
\end{equation*}
\begin{equation*}
\begin{split}
&\frac{1}{2\pi i}\oint_{S^1} \nu g(w)^{2\mu}f(w)^{-2\nu-1}f'(w) dw\\
=&\frac{1}{2\pi i}\oint_{S^1}\left(-\sum_{n=1}^{\infty} nt_{-n} f(w)^{-n} +t_0-\sum_{n=1}^{\infty} v_{-n} f(w)^n\right)g(w)^{\mu}f(w)^{-\nu-1}df(w)\\
=&\frac{1}{\nu}\left(2 \sum_{n=1}^{\infty} nt_{-n}v_{-n} +t_0^2\right).
\end{split}
\end{equation*}Therefore,
\begin{equation*}
\log\mathfrak{T}=-\frac{1}{8}\left(\frac{1}{\mu}+\frac{1}{\nu}\right)t_0^2 +\frac{1}{2}t_0v_0+\frac{1}{2}\sum_{n=1}^{\infty} \left(1-\frac{1}{2\mu}\right)t_nv_n+\frac{1}{2}\sum_{n=1}^{\infty}\left(1-\frac{1}{2\nu}\right)t_{-n}v_{-n}.
\end{equation*}On the other hand, one can show by integration by parts that
\begin{equation*}
\begin{split}
\frac{1}{2\pi i} \oint_{S^1}\mu g(w)^{2\mu-1}f(w)^{-2\nu}g'(w)dw=\frac{1}{2\pi i}\oint_{S^1} \nu g(w)^{2\mu}f(w)^{-2\nu-1}f'(w) dw.
\end{split}
\end{equation*}This gives the nontrivial identity
\begin{equation*}
2\nu\sum_{n=1}^{\infty} nt_nv_n +\nu t_0^2 = 2\mu \sum_{n=1}^{\infty} nt_{-n}v_{-n} +\mu t_0^2.
\end{equation*}

When $\mu=\nu\in \mathbb{N}$, the function
\begin{equation*}
\mathcal{U}(z,\bar{z})=\mathcal{H}(z, \bar{z}^{-1})=|z|^{2\mu}
\end{equation*}is real-valued and regular at $z=0$. By setting $$f(w)=\frac{1}{\overline{g(1/\bar{w})}}$$in \eqref{eq6_17_3} and \eqref{eq6_17_4}, we find that restricted to the subspace $\Sigma$,
\begin{equation*}\begin{split}
t_0=&\frac{\mu}{2\pi i}\oint_{\mathcal{C}_1} z^{\mu-1}\bar{z}^{\mu}dz=\frac{\mu^2}{\pi}\iint_{\Omega_1^+}|z|^{2\mu-2}d^2z=\frac{-\mu}{2\pi i}\oint_{\mathcal{C}_1} z^{\mu}\bar{z}^{\mu-1}d\bar{z},\\
t_n=&\frac{\mu}{2\pi i n}\oint_{\mathcal{C}_1}z^{\mu-n-1}\bar{z}^{\mu}dz =-\frac{\mu^2}{\pi n}\iint_{\Omega_1^-}z^{-n}|z|^{2\mu-2}d^2z=-\overline{\frac{ \mu}{2\pi i n}\oint_{\mathcal{C}_1}z^{\mu}\bar{z}^{\mu-n-1}d\bar{z}}=-\bar{t}_{-n},\\
v_n=&\frac{\mu}{2\pi i }\oint_{\mathcal{C}_1}z^{\mu+n-1}\bar{z}^{\mu}dz =\frac{\mu^2}{\pi}\iint_{\Omega_1^+}z^n |z|^{2\mu-2}d^2z=-\overline{\frac{ \mu}{2\pi i }\oint_{\mathcal{C}_1}z^{\mu}\bar{z}^{\mu+n-1}d\bar{z}}=-\bar{v}_{-n},\\
v_0=&\frac{\mu}{2\pi i}\oint_{\mathcal{C}_1}\left(z^{\mu-1}\bar{z}^{\mu}\log z dz-z^{\mu}\bar{z}^{\mu-1}\log\bar{z}d\bar{z}\right)=\frac{\mu^2}{\pi}\iint_{\Omega_1^+}|z|^{2\mu-2}\log|z|^2d^2z.
\end{split}\end{equation*}Moreover, the restriction of the tau function $\mathfrak{T}$   to $\Sigma$ is equal to
\begin{equation*}
\log\mathfrak{T}=-\frac{t_0^2}{4\mu}+\frac{t_0v_0}{2}+\frac{1}{2}\sum_{n=1}^{\infty}\left(1-\frac{1}{2\mu}\right)\left(t_nv_n+\bar{t}_n\bar{v}_n\right),
\end{equation*}agreeing with the result in \cite{31}.

\end{document}